\documentclass[12pt,a4paper]{article}
\usepackage[margin=1in]{geometry}
\usepackage{graphicx}
\usepackage{subcaption}
\usepackage{amsmath}
\usepackage{amsfonts}
\usepackage{tikz}
\usepackage{float}
\usepackage{authblk}
\usepackage{multirow}
\usepackage{booktabs}
\usetikzlibrary{positioning}
\usepackage{hyperref}
\usepackage{comment}

\usepackage[
    backend=biber,
    style=apa
]{biblatex}

\title{\textbf{A Bayesian Updating Framework for Long-term Multi-Environment Trial Data in Plant Breeding}}
\author{Stephan Bark, Waqas Ahmed Malik, Maryna Prus, Hans-Peter Piepho and Volker Schmid}
\date{}

\begin{document}

\maketitle

\noindent\textbf{Abstract:} In variety testing, multi-environment trials (MET) are essential for evaluating the genotypic performance of crop plants. A persistent challenge in the statistical analysis of MET data is the estimation of variance components, which are often still inaccurately estimated or shrunk to exactly zero when using residual (restricted) maximum likelihood (REML) approaches. At the same time, institutions conducting MET typically possess extensive historical data that can, in principle, be leveraged to improve variance component estimation. However, these data are rarely incorporated sufficiently. The purpose of this paper is to address this gap by proposing a Bayesian framework that systematically integrates historical information to stabilize variance component estimation and better quantify uncertainty. \\
Our Bayesian linear mixed model (BLMM) reformulation uses priors and Markov chain Monte Carlo (MCMC) methods to maintain the variance components as positive, yielding more realistic distributional estimates. Furthermore, our model incorporates historical prior information by managing MET data in successive historical data windows. Variance component prior and posterior distributions are shown to be conjugate and belong to the inverse gamma and inverse Wishart families. While Bayesian methodology is increasingly being used for analyzing MET data, to the best of our knowledge, this study comprises one of the first serious attempts to objectively inform priors in the context of MET data. This refers to the proposed Bayesian updating approach. To demonstrate the framework, we consider an application where posterior variance component samples are plugged into an A-optimality experimental design criterion to determine the average optimal allocations of trials to agro-ecological zones in a sub-divided target population of environments (TPE).

\vspace{0.5cm}

\noindent\textbf{Keywords:} LMM, linear mixed models; BLUP, best linear unbiased prediction; BLMM, Bayesian linear mixed model; ESS, effective sample size; FA, factor-analytic; HMC, Hamiltonian Monte Carlo; INLA, integrated nested Laplace approximation; VI, variational inference; MET, multi-environment trials; MLE, maximum likelihood estimates; MCMC, Markov chain Monte Carlo; MSE, mean squared error; NUTS, No-U-Turn Sampler; RCBD, randomized complete block design; REML, residual (restricted) maximum likelihood; TPE, target population of environments; US, unstructured; VCU, value for cultivation and use.

\newpage

\section{Introduction}

Multi-environment trials (MET) for plant breeding play a central role in modern variety testing, as they provide the empirical basis for assessing genotype performance across diverse environments (\cite{PiephoWilliams2024, SmithCullisThompson2005}). Their importance extends beyond simple performance evaluation: MET are routinely employed in value for cultivation and use (VCU) variety testing, typically conducted in 2–3 year cycles and in long-term stability trials that enable the disentanglement of genetic from non-genetic trends over time (\cite{laidig2014genetic, Rahman2023}). These applications highlight the dual function of MET as both a short-term genotype evaluation tool and a long-term monitoring system for genetic progress. They also play a key role in plant breeding programs, where newly developed candidate lines need to be evaluated across diverse target populations of environments (TPE). \\
The statistical backbone of MET analysis is traditionally known to be the linear mixed model (LMM), with variance components point estimated via residual (restricted) maximum likelihood (REML) (\cite{PattersonThompson1971, gilmour1997}). While REML has become the standard approach, several well documented issues remain. First, variance component estimates may be shrunk to exactly zero when signals are weak (\cite{frey2024analyze}), implicitly dropping random effects from the model and thereby reducing accuracy of the analysis(\cite{Studnicki2024}). Second, although the incorporation of more historical data could, in principle, improve the precision of variance component estimation, the increasing complexity of random effect structures often leads to convergence difficulties as the amount of data grows (\cite{stranden2024computationally}). These limitations motivate the search for alternative inferential frameworks that can better exploit historical information while maintaining computational stability. \\
A promising alternative is the adoption of a fully Bayesian approach, where variance components of random effects are treated as random variables with prior distributions, and posterior inference is obtained using Markov chain Monte Carlo (MCMC) methods. When sufficient historical data is available, Bayesian methods can provide more robust and stable distributional estimates of variance components, particularly if priors are carefully chosen (\cite{oliveira2016, silva2019, nuvunga2019, daSilva2023ammi}). Such approaches provide a principled
way to quantify uncertainty in variance component estimation. \\
It is surprising that for variety testing, however, hardly any attention seems to have been paid to leveraging historical databases to objectively inform variance component priors across successive historical windows. Addressing this gap, the present work develops a Bayesian framework for analyzing large historical MET databases, in which variance component priors are regularly updated as new data from current testing cycles become available. Following the terminology of Section 5.5 in the textbook of \textcite{sorensen2002likelihood}, this iterative incorporation of historical information is referred to as a Bayesian updating approach. In addition to the prior information in the variance components justified by real historical data, we consider this approach to be more robust than the common practice of feeding a large MET dataset into the Bayesian linear mixed model (BLMM) all at once. \\
We expect our framework to deliver a basis for a new generation of BLMM for MET that is more meaningful in terms of the systematic incorporation of prior knowledge and is potentially more efficient and more stable than other current models, which still do not benefit from data usage in historical windows. Note that our proposed method can be applied in two closely related contexts: (i) breeding programs and (ii) variety testing. While these areas differ in their objectives and operational structures, they share many methodological commonalities in the analysis of MET data, making our framework relevant for both. \\ For demonstration purposes, we consider a long-term dataset based on VCU trials for rice in Bangladesh, described in \textcite{Rahman2023}. This data clearly falls into category (ii), where we went through an implementation example. First, we specified and implemented our Bayesian updating framework. It contains four climatic zones of Bangladesh. Second, we illustrated Bayesian updating, considering an optimal design problem in MET for a sub-divided TPE, as discussed by \textcite{Prus2024}.

\section{Materials and Methods}

\subsection{Data}
\label{sec:data}
The experimental design of the data is described in \textcite{Rahman2023}. 41 varieties released in the winter season and 45 varieties released in the monsoon season were examined for the years from 2001-2002 to 2020-2021 in Bangladesh. Their grain yield was measured in tons per hectare (t/ha). The trials were carried out as a randomized complete block design (RCBD) with three replicates at eight different locations in Bangladesh. The eight locations can be assigned to four climatic zones of Bangladesh. While some varieties were released in the years from 1970 to 2001, other varieties were added to the dataset after their release. Every year, a new randomization is carried out for the current set of genotypes. Based on that, it becomes clear that the genotypes in each replicate nested in the location will not be allocated to the same individual plot each year by design. This becomes important when modeling the data. \\
It should be mentioned that we labeled the varieties with respect to three different genotype groups: 'Short', 'Medium', and 'Long'. The Winter season genotype group 'Medium' data is sufficient for the purposes of this work and was primarily analyzed within its context. 

\subsection{Statistical analysis}
\begin{figure}[ht]
    \centering
    \includegraphics[width=1\textwidth]{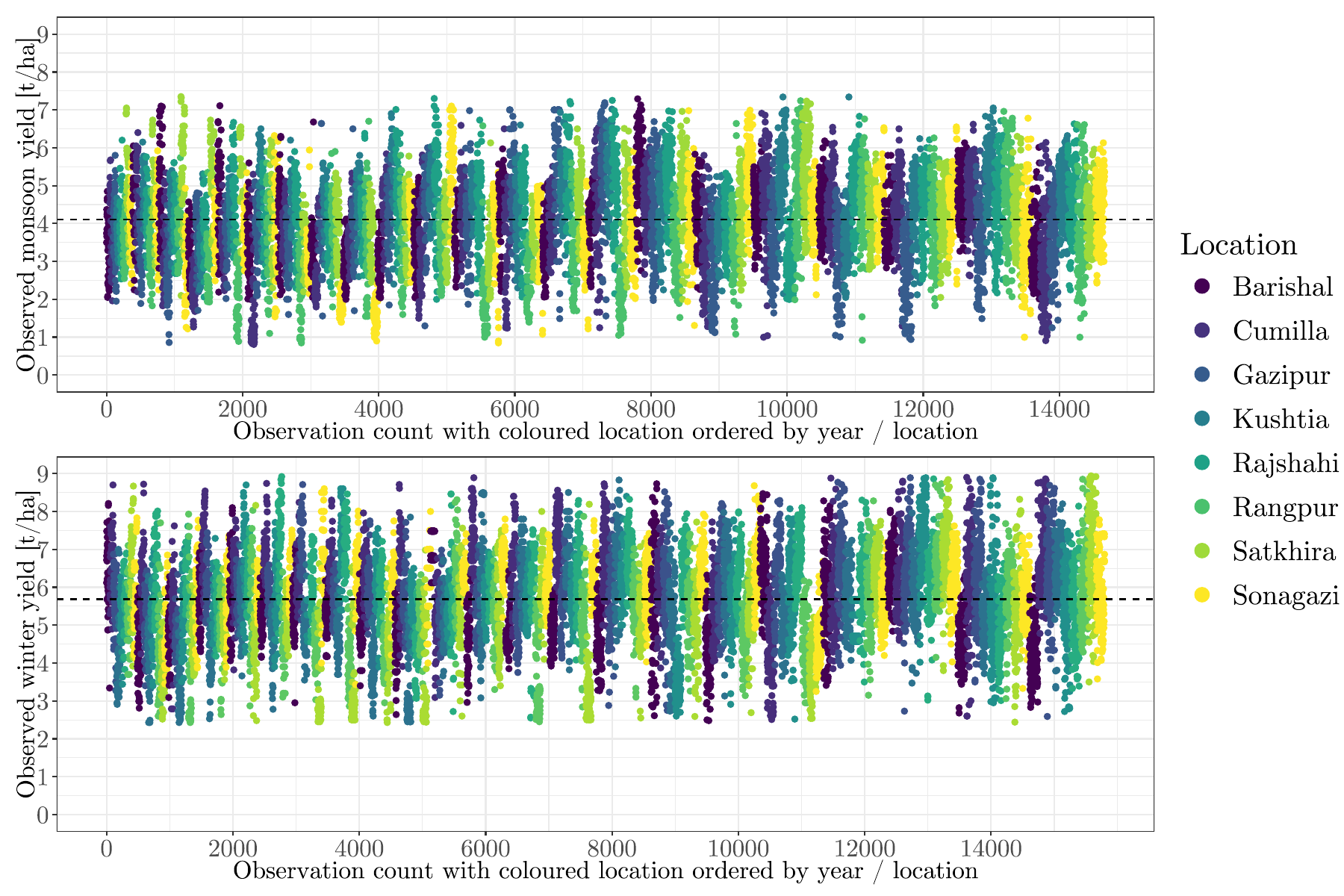}
    \caption[Ordered Grain Yield Plot]{Observed yield in tons per hectare as the response variable of interest plotted and sorted by environment (year-location combination) from \textcite{Rahman2023}; above for monsoon genotypes; below for Winter genotypes.}
    \label{fig:observed:yield}
\end{figure}

In the descriptive plot of Figure \ref{fig:observed:yield}, the variance heterogeneity of the residual error between the different year-location combinations stands out. We accounted for this heterogeneity when modeling this data. In addition, based on discussions with colleagues in Bangladesh, a routine outlier adjustment was carried out in which the detected outliers were replaced with corresponding imputed values. \\
The resulting dataset is almost complete and has only two missing values in the Winter yield data. The dataset is highly imbalanced regarding the different years, as new genotypes are added to the experiment over the years.

\subsubsection{Frequentistic linear mixed model}
\label{sec:methods:MM:specification}
Justified by the underlying experimental design in the data described in \textcite{Rahman2023}, we modeled locations as nested within years (Chapter 8 of \cite{Patterson1997}). All location-specific effects that do not depend on year-specific random effects were dropped. Note that the model was constructed for individual genotype groups 'Short', 'Medium', and 'Long', nested in each of the two seasons, which correspond to monsoon and winter. The corresponding number of genotypes in the \(q\)-th group is called \(M_{\text{Monsoon}; q}\) and \(M_{\text{Winter}; q}\) with $q$ = 1, 2, 3 for 'Short', 'Medium', and 'Long', respectively. The \(v\)-th replicate of the \(g\)-th genotype in the \(h\)-th year at the \(z\)-th zone and the \(c\)-th location is modeled as the random variable
\begin{equation} \label{MM:specification:formular}
y_{zchgv} = \mu + \zeta_z + \eta_h + \beta_{zh} + \delta_{zch} + b_{zchv} + \alpha_{zg} + \omega_{hg} + \tau_{zhg} + \varphi_{zchg} + \varepsilon_{zchgv},
\end{equation}
for \(z = 1, \dots, 4\), \(c = 1, \dots, 8\), \(h = 1, \dots, 22\), \(g = 1, \dots, M_{\text{Monsoon}; q}\), and \(v = 1, 2, 3\) in the Monsoon season data. For the Winter season data, the indices are given by \(z = 1, \dots, 4\), \(c = 1, \dots, 9\), \(h = 1, \dots, 22\), \(g = 1, \dots, M_{\text{Winter}; q}\), and \(v = 1, 2, 3\). Note that only the data corresponding to the Winter season genotype group 'Medium' was actively used for the analysis in this study. \\
The response variable of interest \(y_{zchgv}\) represents the yield in tons per hectare. The global intercept is denoted by \(\mu\). The term \(\zeta_z\) accounts for the fixed effect of the \(z\)-th zone. The random effect of the \(h\)-th year is given by \(\eta_h\), which is assumed to follow an independent and identically distributed (i.i.d.) normal distribution, \(\eta_h \stackrel{i.i.d.}{\sim} N(0, \sigma_{\eta}^2)\). The random effect of the \(z\)-th zone and \(h\)-th year is denoted by \(\beta_{zh}\), following \(\beta_{zh} \stackrel{i.i.d.}{\sim} N(0, \sigma_{\beta}^2)\). Similarly, \(\delta_{zch}\) represents the random effect of the \(z\)-th zone, \(c\)-th location, and \(h\)-th year, assuming \(\delta_{zch} \stackrel{i.i.d.}{\sim} N(0, \sigma_{\delta}^2)\). The term \(b_{zchv}\) captures the random effect of the \(v\)-th replicate nested within the \(\delta_{zch}\) effect, with \(b_{zchv} \stackrel{i.i.d.}{\sim} N(0, \sigma_b^2)\). The random effect of the \(z\)-th zone and the \(g\)-th genotype is represented by \(\alpha_{zg}\), following \(\alpha_{zg} \stackrel{i.i.d.}{\sim} N(0, \mathbf{I}_{\alpha} \otimes \mathbf{\Sigma}_{\alpha})\), where \(\otimes\) denotes the Kronecker product. This results in \(M_{\text{Monsoon}}\) identical \(4 \times 4\) covariance matrices \(\mathbf{\Sigma}_{\alpha}\) along the diagonal, each corresponding to one genotype and reflecting the unstructured (US) assumption of the zones in the genotype-zone random effect. The random effect of the \(h\)-th year and the \(g\)-th genotype is given by \(\omega_{hg}\), assuming \(\omega_{hg} \stackrel{i.i.d.}{\sim} N(0, \sigma_{\omega}^2)\). The random effect of the \(z\)-th zone nested within \(\omega_{hg}\) is denoted as \(\tau_{zhg}\) with \(\tau_{zhg} \stackrel{i.i.d.}{\sim} N(0, \sigma_{\tau}^2)\), while \(\varphi_{zchg}\) represents the random effect of the \(c\)-th location nested within \(\tau_{zhg}\), following \(\varphi_{zchg} \stackrel{i.i.d.}{\sim} N(0, \sigma_{\varphi}^2)\). Finally, the residual error \(\varepsilon_{zchgv}\) is assumed to be independently (ind.) distributed for each \(c\)-th location and \(h\)-th year combination, such that \(\varepsilon_{zchgv} \stackrel{ind.}{\sim} N(0, \sigma_{\varepsilon; ch}^2)\). \\
For ease of reference, the variance components of the model will be represented as the following short forms: \( \sigma_{\eta}^2 \) as var\_year, \( \sigma_{\beta}^2 \) as var\_zone\_year, \( \sigma_{\delta}^2 \) as var\_zone\_loc\_year, \( \sigma_b^2 \) as var\_zone\_loc\_rep\_year, \( \sigma_{\alpha}^2 \) as var\_gen\_zone, \( \sigma_{\omega}^2 \) as var\_gen\_year, \( \sigma_{\tau}^2 \) as var\_gen\_zone\_year, \( \sigma_{\varphi}^2 \) as var\_gen\_zone\_loc\_year, and \( \sigma_{\varepsilon; ch}^2 \) as var\_resid\_env[\(\cdot\)], where, e.g., var\_resid\_env[1] corresponds to the first environment in the dataset. The mean of all environmental residual variance components will be assigned to env\_mean\_var\_resid. Also, var\_gen\_zone will be extended to Genotype \(\times\) Zone [\(\cdot\),\(\cdot\)], where, e.g., Genotype \(\times\) Zone [1,2] corresponds to the element in the first row and the second column of a matrix-valued variance-covariance structure corresponding to the genotype-zone random-effect variance assumptions.

\subsubsection{Bayesian linear mixed model reformulation}
\label{sec:bayesmod:bayes:cond}
The LMM described in the previous section, involving an US structural assumption with respect to the zones involved in the genotype-zone effect, will be reformulated and extended by the addition of another layer in the hierarchical model structure corresponding to prior distributional assumptions on the random effect variance components. For simplicity and the property of conjugacy with respect to prior and posterior, the inverse gamma distribution is chosen for scalar-valued variance components, and the inverse Wishart distribution for the matrix-valued variance-covariance component. In-depth derivations can be found in the Appendix Section \ref{app:bayesmod:bayes:cond}. In the following, the joint posterior of the general fixed intercept \(\mu\), all other fixed effects \(\zeta_z\), all random effects \(\eta_h, \beta_{zh}, \delta_{zch}, b_{zchv}, \alpha_{zg}, \omega_{hg}, \tau_{zhg}, \varphi_{zchg}\), and all their corresponding variance components, as well as the residual error variance components, will be stated.

\subsubsection{Bayesian updating sampling algorithm}
Following the Gibbs sampling methodology, we accessed the joint posterior by drawing marginal parameter samples from its full conditionals. For simplicity in notation, all model parameters are collected in the placeholder parameter vector \(\boldsymbol{\theta}\). With the response observation \(y_{zchgv}\) all collected into the response vector \(\mathbf{y}\), Bayes' theorem can be applied in its general form, given as:
\begin{equation}
    p(\boldsymbol{\theta} \mid \mathbf{y}) = \frac{\mathcal{L}(\mathbf{y} \mid \boldsymbol{\theta}) p(\boldsymbol{\theta})}{p(\mathbf{y})} \stackrel{wrt. \boldsymbol{\theta}}{\propto} \mathcal{L}(\mathbf{y} \mid \boldsymbol{\theta}) p(\boldsymbol{\theta}),
\end{equation}
where \(\mathbf{p}\left(\boldsymbol{\theta}\mid\mathbf{y}\right)\) is the joint posterior distribution, \(\mathcal{L}\left(\mathbf{y}\mid\boldsymbol{\theta}\right)\) is the likelihood function, and \(\mathbf{p}\left(\boldsymbol{\theta}\right)\) is the joint prior distribution. Note that \(\mathbf{p}\left(\mathbf{y}\right)\) refers to the marginal data distribution, which is mostly not traceable in practice. One of the main challenges in Bayesian inference is to make \(\mathbf{p}\left(\boldsymbol{\theta}\mid\mathbf{y}\right)\) accessible without knowing \(\mathbf{p}\left(\mathbf{y}\right)\). MCMC simulation lays the theoretical foundation for the idea of sampling from \(\mathbf{p}\left(\boldsymbol{\theta}\mid\mathbf{y}\right)\) (\cite{Gelman2013} chapters 11 and 12) and will be considered in this study. The Gibbs sampler provides an efficient sampling algorithm that directly samples from the full conditional posterior distributions of all parameters \(\mathbf{\theta}_{1},\ldots,\mathbf{\theta}_{D}\) and generates samples of all corresponding marginal parameter distributions \(\mathbf{p}\left(\mathbf{\theta}_{d}\mid\mathbf{y}\right)\). \\
An in-depth specification of the developed sampling algorithm in the specific given model context is provided in the Appendix Section \ref{app:bayesmod:bayes:sampling}. The general principle refers to a given multi-year MET dataset. The data is divided into \({L}\) multi-year historical windows. We propose to chose \({L}\) between 3 and 6, within a minimum of 2 years per window. Based on this, the empirical information contained in the marginal posterior distributions of the LMM variance components from each historical window is carried forward to the next. In the subsequent window, these posterior distributions will serve as the current historical prior information. After achieving convergence in the current window, we can sample successively from the posterior distribution for each multi-year window dataset \(\mathbf{y}_l\) with \(l = 1, \dots, L\): 
\begin{enumerate}
    \item Initialize \( \boldsymbol{\theta}_{(l)}^{(0)} = ((\theta_1)_{(l)}^{(0)}, \dots, (\theta_D)_{(l)}^{(0)})^\top \) as draws from the priors.
    \vspace{0.2cm}
    \item For iteration \( t = 1, \dots, T \):
    \begin{itemize}
    \vspace{0.1cm}
       \item[(1)] Sample \( (\theta_1)_{(l)}^{(t)} \sim \ \ \ p_{(l)}(\theta_1 | (\theta_2)_{(l)}^{(t-1)}, \dots, (\theta_D)_{(l)}^{(t-1)}, \mathbf{y}_{(l)}) \)
       \vspace{0.1cm}
       \item[(2)] Sample \( (\theta_2)_{(l)}^{(t)} \sim \ \ \ p_{(l)}(\theta_2 | (\theta_1)_{(l)}^{(t)}, (\theta_3)_{(l)}^{(t-1)}, \dots, (\theta_D)_{(l)}^{(t-1)}, \mathbf{y}_{(l)}) \)
       \item[] \( \vdots \)
       \item[(D)] Sample \( (\theta_D)_{(l)}^{(t)} \sim \underbrace{p_{(l)}(\theta_D | (\theta_1)_{(l)}^{(t)}, \dots, (\theta_{D-1})_{(l)}^{(t)}, \mathbf{y}_{(l)})}_{\text{Full cond. posterior implicitly depends on} \ \mathbf{y}_{(1)}, \dots, \mathbf{y}_{(l-1)}} \)
    \end{itemize}
    \vspace{0.2cm}
    \item Compute posterior distributional parameters of the variance components for their priors of the next window \(l+1\) via maximum likelihood estimation using the drawn marginal posterior samples of the current window \(l\).
\end{enumerate}

The Bayesian updating sampling algorithm was implemented using the \texttt{rstan} package (\cite{Stan2022}) in \texttt{R}, employing the No-U-Turn Sampler (NUTS), a variant of Hamiltonian Monte Carlo (HMC), for efficient posterior sampling. To ensure convergence, the first window, containing eight years of historical data, was sampled with 37,500 iterations, including 30,000 burn-in iterations. The subsequent four windows contain five years in the second window and three years in the remaining windows, using 27,500 iterations, including 20,000 burn-in iterations. Although this seems to be much less than is commonly used (\cite{przystalski2020, przystalski2023}), convergence was achieved. Convergence was evaluated using a set of Stan diagnostic measures, namely the Gelman-Rubin scale reduction factor, also known as R-hat, effective sample size (ESS), and Geweke z-score, following standard practices as stated in Section 1.12 in the textbook by \textcite{congdon2020bayesian}. We provide a reproducible HTML report on our webpage at \href{https://github.com/StephanBark/Bayesian_Updating_with_Optimal_Design_2026}{GitHub}. \\

\newpage

Four chains were run in parallel with a thinning factor of two to reduce autocorrelation within the chains. Overall, this leads to a posterior sample size of 10000, considering the \texttt{rstan} function \texttt{extract} that permutes and merges all draws of all chains after burn-in. The adaptation delta was set to 0.95 for the first window and 0.9 for the remaining windows, balancing exploration and convergence. The maximum tree depth was set to 12 and 10, respectively, and the step size was fixed at 0.1. A fixed random seed ensured reproducibility. All computations were performed on a Windows system with 32 gigabytes of random-access memory and 6 central processing unit cores, which was sufficient to accommodate the memory demands of the model and sampling process based on the MET dataset described in \textcite{Rahman2023} on the winter rice "Medium" genotype group containing 5318 observations. \\
Initial prior parameters were chosen to reflect weakly informative beliefs while ensuring computational stability. Inverse gamma priors were used for the variance components, with shape parameters of five and scale parameters of one each. The genotype-zone-related covariance matrix of zones \(\mathbf{\Sigma}_Z\) was assigned an inverse Wishart prior with \(\nu = 10\) degrees of freedom and a scale matrix \(\mathbf{S}\) with diagonal entries set to 1 and off-diagonal entries set to 0.9. These choices reflect moderate prior certainty and allow for flexible learning across windows.

\subsubsection{A-Optimal design application}
\label{method:A:optimal:design}
A-optimality is a concept in the context of the design of experiments focused on minimizing the trace of the inverse of an information matrix (\cite{John1995}). In the LMM context in plant breeding, this corresponds to the mean squared error (MSE) matrix of the best linear unbiased prediction (BLUP) of random genotypic effects or their linear contrasts (\cite{Prus2024}). MET can be designed so that genotypes BLUP have the smallest possible average variance given a fixed amount of experimental resources, such as the total number of trials. For the 22 year MET data from \textcite{Rahman2023}, optimality refers to the optimal number of trials allocated to each of the four climatic zones specified for Bangladesh to enhance the estimation of genotype effects and their pairwise contrasts in a LMM framework. In this study, the optimal design refers to the approximate design \(\boldsymbol{\xi^{\star}_a}\) containing the optimal zone weights \(w_1^{\star}\), \(w_2^{\star}\), \(w_3^{\star}\), and \(w_4^{\star}\), which sum up to one. We build our Bayesian framework based on the A-optimality design criterion \(\Phi^{\star}_A(\boldsymbol{\xi_a})\) for LMM on MET data by \textcite{Prus2024}. Formal definitions are provided in the Appendix Section \ref{app:opti:design}, which also involves the evaluation criteria \(Eff_{a}\) and \( MSE_{\text{Tr}} \) used in our result Tables \ref{tab:design:freq:US:with:year} and \ref{tab:bayes:design:with:year}. \\
To illustrate our Bayesian updating modeling approach, we follow up on \(\boldsymbol{\xi^{\star}_a}\) with LMMs. As before by \textcite{Prus2024}, our practical objective is to determine how a fixed number of trials should be distributed across different climatic zones so that genotype performance can be estimated as precisely as possible for each of the zones. However, in this context, uncertainty in the variance components poses a key challenge, as the optimal design criteria \(\Phi^{\star}_A(\boldsymbol{\xi_a})\) depend directly on these parameters. Ignoring this uncertainty can lead to bias in the optimal design calculation, especially since its point estimates are less robust than when parameter uncertainty is taken into account. \\

\newpage

To address this issue, we integrated Bayesian inference into the optimal design methodology of \textcite{Prus2024}. Replacing the REML approach of the MET data model with our recommended Bayesian Updating framework, the posterior distributional uncertainty measures of variance components are directly linked to their corresponding optimal design criteria uncertainty. In particular, the A-optimality criterion of \textcite{Prus2024} can be extended to incorporate posterior samples, thereby allowing not only the optimal allocation of trials across zones to be evaluated, but also the uncertainty associated with these allocations. \\
To generate these samples of the approximate design weight \(w_1\) up to \(w_4\), the inverse gamma and inverse Wishart distributions, approximated from the variance component posterior samples of the last historical window, are used to draw 100 sets of variance components corresponding to our BLMM. For each variance component set, its approximate design \(\boldsymbol{\xi^{\star}_a}\) was calculated. Accordingly, \(\boldsymbol{\bar{\xi}^{\star}_a}\) related to the weight means \(\bar{w}_1^{\star}\), \(\bar{w}_2^{\star}\), \(\bar{w}_3^{\star}\), and \(\bar{w}_4^{\star}\) of the weights \(w_1^{\star}\), \(w_2^{\star}\), \(w_3^{\star}\), and \(w_4^{\star}\) with respect to our 100 optimal design samples. \\
The computation of the A-optimal design was performed using the `OptimalDesign` R-package with its \texttt{od\_MISOCP} function (\cite{Harman2022}). Note that the design criterion, as described above, is reformulated to fit as an input for the \texttt{od\_MISOCP} function. \textcite{HARMAN2018} provide the required mathematical derivations for that purpose.

\section{Results}

\begin{figure}[ht]
    \includegraphics[width=1\textwidth]{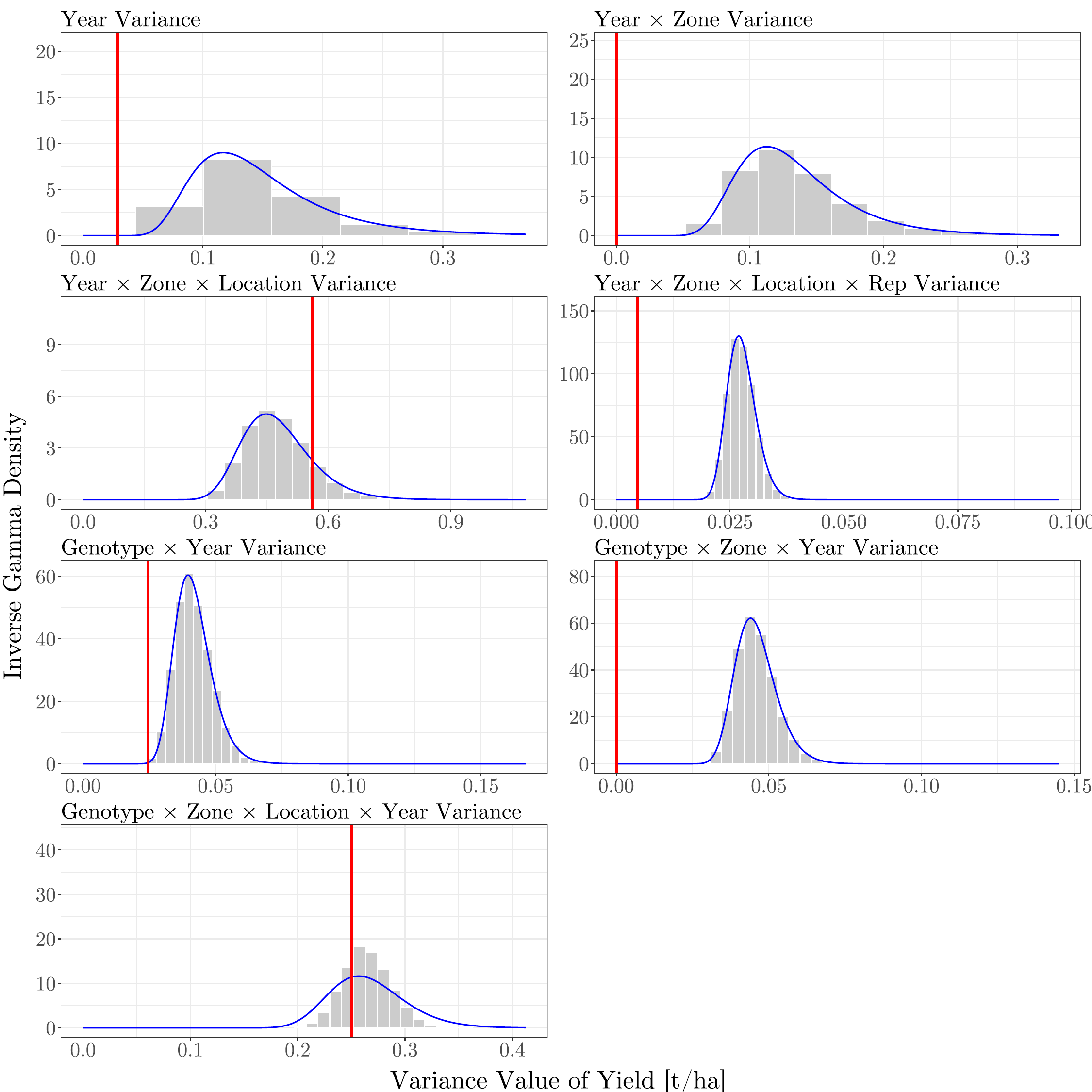}
    \caption[Posterior Histograms of Other Variance Components of BLMM after the Fifth Sampled Window]{Histograms of the drawn MCMC posterior samples after the fifth multi-year historical window for all random effect variance components expect of the Sigma\_gen\_zone matrix (in grey) corresponding to the 10000 posterior sampled values of MCMC chain one. Maximum likelihood estimated posterior density under inverse gamma approximation assumption referring to the current marginal samples of variance components (in blue). The frequentist REML point estimates computed with the ASReml package (\cite{Butler2023}) in the R software (in red). On the y-axis is the density function of the inverse gamma, on the x-axis is the values of the drawn variance components. Note that in all histograms the breaks are fixed to 15. Wider bars hint to a wider spread of the values plotted along the x-axis.}
\label{fig:post:histo:AllOther:cycle5:corr09}
\end{figure}

\begin{figure}[htp]
    \centering
    \begin{subfigure}[b]{1\textwidth}
        \includegraphics[width=\textwidth]{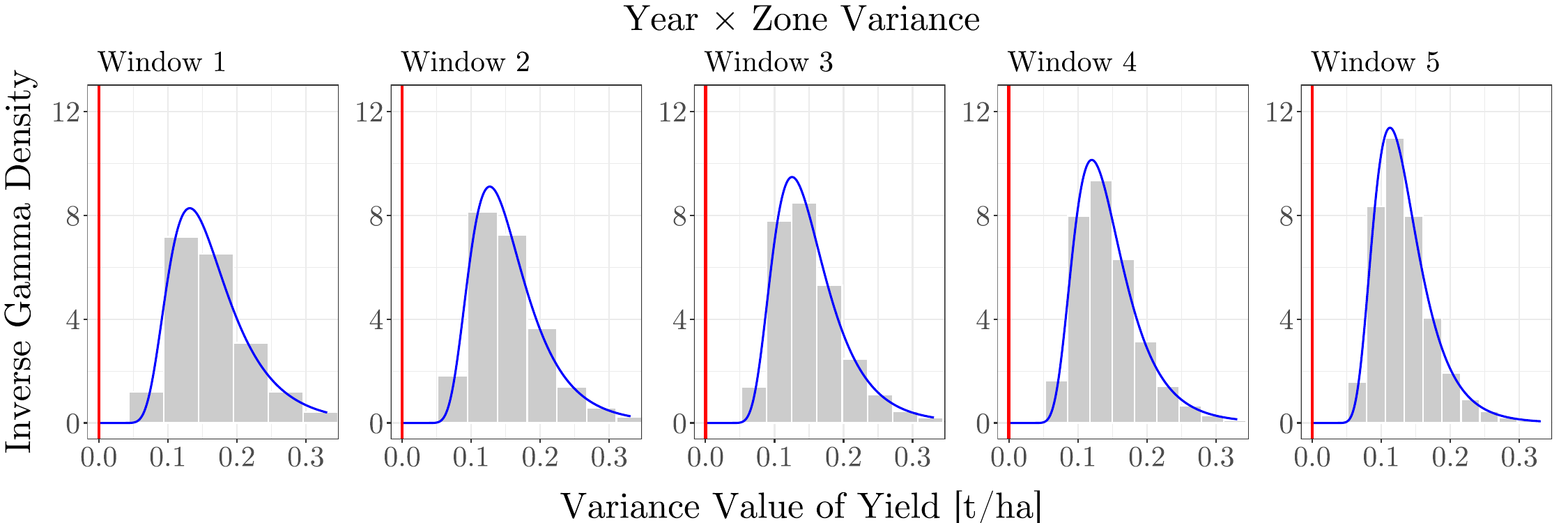}
        \label{fig:post:histo:ALL:WINDOW:ZONE:YEAR:corr09}
    \end{subfigure}
    \hfill
    \begin{subfigure}[b]{1\textwidth}
        \includegraphics[width=\textwidth]{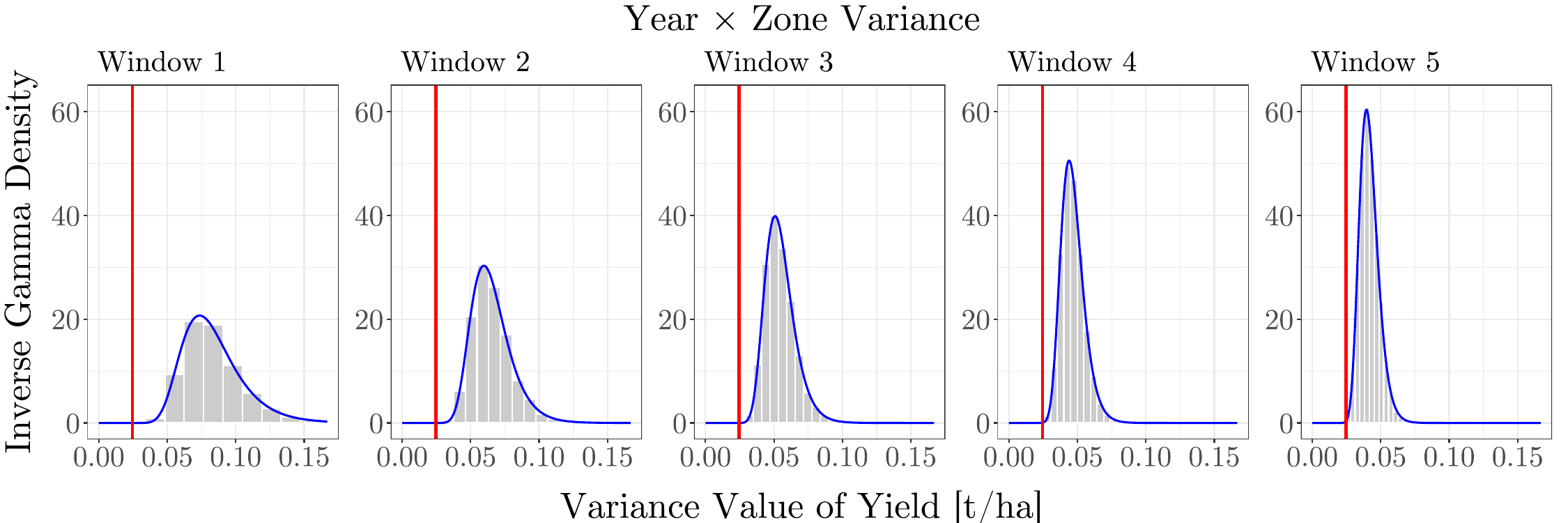}
        \label{fig:post:histo:ALL:WINDOW:GEN:YEAR:corr09}
    \end{subfigure}
    \caption[Posterior Histograms of the var\_gen\_zone and var\_zone\_year components of BLMM of the all successive sampled windows]{Histograms of the drawn MCMC posterior samples of all successive sampled multi-year historical window for the var\_gen\_zone and var\_zone\_year components expect (in grey) corresponding to the 10000 posterior sampled values per window of MCMC chain one out of four. Maximum likelihood estimated posterior density under inverse gamma approximation assumption referring to the current marginal samples of variance components (in blue). The frequentist REML point estimates computed with the ASReml package (\cite{Butler2023}) in the R software (in red). On the y-axis is the density function of the inverse gamma, on the x-axis is the values of the drawn variance components. Note that in all histograms the breaks are fixed to 15. Wider bars hint to a wider spread of the values plotted along the x-axis.}
    \label{fig:post:histo:ALL:WINDOW:corr09}
\end{figure}

\begin{figure}[ht]
    \includegraphics[width=0.9\textwidth]{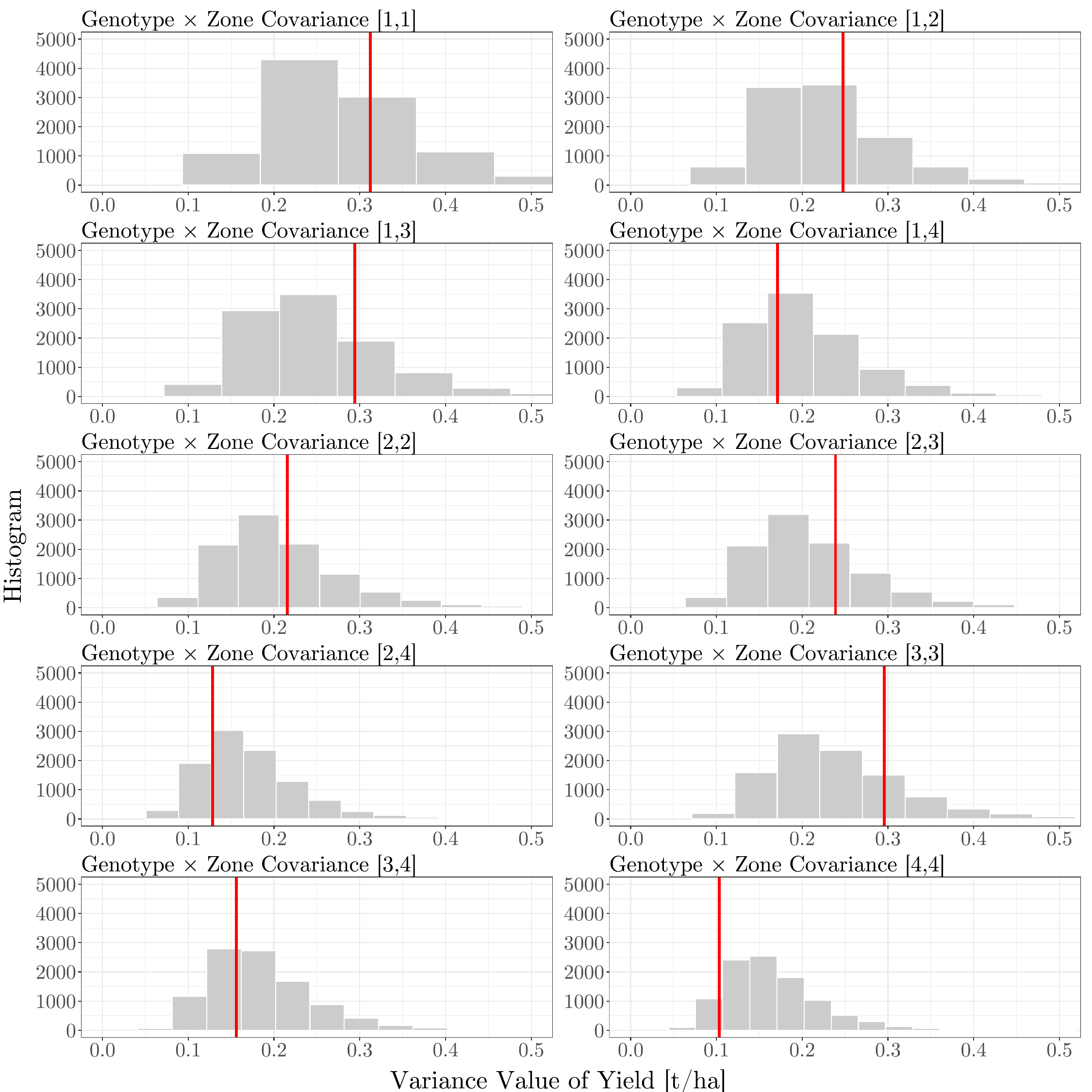}
    \caption[Posterior Histograms of Sigma\_gen\_zone Component of BLMM after the Fifth Sampled Window]{Histograms of the drawn MCMC posterior samples for variance-covariance parameters of the Sigma\_gen\_zone matrix after the fifth multi-year historical window (in grey) corresponding to the 10000 posterior sampled values of MCMC chain one. The frequentist REML point estimates computed with the ASReml package (\cite{Butler2023}) in the R software (in red). On the y-axis is the MCMC posterior sample count, on the x-axis is the values of the drawn variance components. Note that in all histograms the breaks are fixed to 15. Wider bars hint to a wider spread of the values plotted along the x-axis.}
\label{fig:post:histo:GenZone:cycle5:corr09}
\end{figure}

In the following, the most important posterior sampling results corresponding to the MET data described in \textcite{Rahman2023} analyzed in five historical windows with our specified Bayesian updating algorithm are described and visualized in the Figures \ref{fig:post:histo:AllOther:cycle5:corr09}, \ref{fig:post:histo:ALL:WINDOW:corr09}, and \ref{fig:post:histo:GenZone:cycle5:corr09}. For the intermediate windows, plots can also be generated using the R code, which can be found on the Git website linked in the Appendix Section \ref{supp:mat}. Note that the correlation of the initial CS structure of the scale matrix parameter of the inverse Wishart prior corresponding to Sigma\_gen\_zone was chosen to be 0.9 in the Figures \ref{fig:post:histo:AllOther:cycle5:corr09}, \ref{fig:post:histo:ALL:WINDOW:corr09} and \ref{fig:post:histo:GenZone:cycle5:corr09}.

\clearpage

\clearpage

In Tables \ref{tab:design:freq:US:with:year} and \ref{tab:bayes:design:with:year}, the optimal design results are in the same format as Table 3 of \textcite{Prus2024}. The Tables \ref{tab:design:freq:US:with:year} and \ref{tab:bayes:design:with:year} are based on the REML model and on posterior samples of variance components. Note that the total number of three years is \(H\). The three-year window reflects the typical duration of VCU testing in Germany and other European countries, where new experimental breeding cycles start annually, with trials conducted at multiple locations. Fortunately, the three-year cycles reflect the typical VCU testing system in Bangladesh as well. 

\clearpage
In Table \ref{tab:design:freq:US:with:year}, optimal allocations of trials regarding the approximate designs, as performed by \textcite{Prus2024}, are demonstrated. Averages and their standard deviations are presented for the same optimality criterion based on our recommended Bayesian updating variance component posterior sample results. Note that in some rows of Table \ref{tab:bayes:design:with:year}, the weights \(w_1\) up to \(w_4\) do not exactly add up to one because of the averaging process as described in Section \ref{method:A:optimal:design}. However, the weights of the 100 sampled designs each add exactly up to one individually.
\begin{table}[ht]
\caption[Grid of A-Optimal designs of our LMM involving an US structural assumption with respect to the zones involved in the genotype-zone effect under consideration of multiple years]{Optimal numbers of locations per zone and efficiency of balanced design compared to optimal designs with respect to the standard A-criterion as well as the MSE matrix trace in case of US variance-covariance structural assumption of Sigma\_gen\_zone in the model for three years \(H\) and different values of the total number locations \(J\).}
\label{tab:design:freq:US:with:year}
\vspace{0.5cm}
\begin{tabular}{|l|l|llll|l|l|}
\cline{1-2} \cline{7-8}
\multicolumn{1}{|c|}{\multirow{2}{*}{\(H\)}} &
\multicolumn{1}{c|}{\multirow{2}{*}{\(J\)}} &
\multicolumn{4}{c|}{\textbf{Approximate design \(\boldsymbol{\xi^{\star}_a}\)}\footnotemark[1]} &
\multicolumn{1}{c|}{\multirow{2}{*}{\textbf{\(Eff_{a}\)}\footnotemark[2]}} &
\multicolumn{1}{c|}{\multirow{2}{*}{\textbf{\(MSE_{\text{Tr}}\)}\footnotemark[3]}} \\ \cline{3-6}
\multicolumn{1}{|c|}{} &
\multicolumn{1}{c|}{} &
\multicolumn{1}{c|}{\(w_1^{\star}\)} &
\multicolumn{1}{c|}{\(w_2^{\star}\)} &
\multicolumn{1}{c|}{\(w_3^{\star}\)} &
\multicolumn{1}{c|}{\(w_4^{\star}\)} &
\multicolumn{1}{c|}{} &
\multicolumn{1}{c|}{} \\ \hline

3 & 10  & \multicolumn{1}{l|}{0.52} & \multicolumn{1}{l|}{0.1}  & \multicolumn{1}{l|}{0.28} & 0.1  & 0.89 & 0.0922 \\ \hline
3 & 20  & \multicolumn{1}{l|}{0.49} & \multicolumn{1}{l|}{0.17} & \multicolumn{1}{l|}{0.29} & 0.05 & 0.9  & 0.0605 \\ \hline
3 & 40  & \multicolumn{1}{l|}{0.44} & \multicolumn{1}{l|}{0.22} & \multicolumn{1}{l|}{0.29} & 0.05 & 0.94 & 0.0412 \\ \hline
3 & 100 & \multicolumn{1}{l|}{0.38} & \multicolumn{1}{l|}{0.22} & \multicolumn{1}{l|}{0.26} & 0.14 & 0.96 & 0.0236 \\ \hline
3 & 200 & \multicolumn{1}{l|}{0.38} & \multicolumn{1}{l|}{0.21} & \multicolumn{1}{l|}{0.24} & 0.18 & 0.95 & 0.0146 \\ \hline

\end{tabular}
\end{table}

\begin{table}[ht]
\caption[Grid of A-Optimal designs of our BLMM involving an US structural assumption with respect to the zones involved in the genotype-zone effect under consideration of multiple years]{Average optimal allocations of trials and standard deviations in brackets of optimal approximate design and efficiency of balanced design compared to optimal designs with respect to the standard A-criterion as well as the MSE matrix trace in case of US variance-covariance structural assumption of Sigma\_gen\_zone in the model for three years \(H\) and different values of the total number locations \(J\).}
\label{tab:bayes:design:with:year}
    \vspace{0.5cm}
    \begin{tabular}{|l|l|llll|l|l|}
    \cline{1-2} \cline{7-8}
    \multicolumn{1}{|c|}{\multirow{2}{*}{\(H\)}} & \multicolumn{1}{c|}{\multirow{2}{*}{\(J\)}} & \multicolumn{4}{c|}{\textbf{Average of approximate design \(\boldsymbol{\bar{\xi}^{\star}_a}\)}\footnotemark[1]}                                                                                & \multicolumn{1}{c|}{\multirow{2}{*}{\textbf{\(\overline{Eff}_{a}\)}\footnotemark[2]}} & \multicolumn{1}{c|}{\multirow{2}{*}{\textbf{\(\overline{MSE}_{\text{Tr}}\)}\footnotemark[3]}} \\ \cline{3-6}
    \multicolumn{1}{|c|}{} & \multicolumn{1}{c|}{} 
    & \multicolumn{1}{c|}{\(\bar{w}_1^{\star}\)} 
    & \multicolumn{1}{c|}{\(\bar{w}_2^{\star}\)} 
    & \multicolumn{1}{c|}{\(\bar{w}_3^{\star}\)} 
    & \multicolumn{1}{c|}{\(\bar{w}_4^{\star}\)} 
    & \multicolumn{1}{c|}{} 
    & \multicolumn{1}{c|}{} \\ \hline

    3 & 10  & \multicolumn{1}{l|}{0.4 (0.07)}  & \multicolumn{1}{l|}{0.18 (0.07)} & \multicolumn{1}{l|}{0.29 (0.08)} & 0.12 (0.04) & 0.93 (0.031) & 0.067 (0.0089) \\ \hline
    3 & 20  & \multicolumn{1}{l|}{0.37 (0.06)} & \multicolumn{1}{l|}{0.21 (0.06)} & \multicolumn{1}{l|}{0.29 (0.06)} & 0.12 (0.06) & 0.94 (0.032) & 0.038 (0.0051) \\ \hline
    3 & 40  & \multicolumn{1}{l|}{0.34 (0.04)} & \multicolumn{1}{l|}{0.22 (0.05)} & \multicolumn{1}{l|}{0.28 (0.04)} & 0.15 (0.05) & 0.95 (0.029) & 0.021 (0.003)  \\ \hline
    3 & 100 & \multicolumn{1}{l|}{0.32 (0.04)} & \multicolumn{1}{l|}{0.23 (0.04)} & \multicolumn{1}{l|}{0.27 (0.04)} & 0.18 (0.04) & 0.96 (0.025) & 0.01 (0.0014)  \\ \hline
    3 & 200 & \multicolumn{1}{l|}{0.31 (0.03)} & \multicolumn{1}{l|}{0.23 (0.04)} & \multicolumn{1}{l|}{0.27 (0.03)} & 0.19 (0.04) & 0.96 (0.025) & 0.005 (8e-04)  \\ \hline

    \end{tabular}
\end{table}
\footnotetext[1]{More details can be found in our Appendix Section \ref{app:opti:design} and the work of \textcite{Prus2024} regarding \(\boldsymbol{\xi^{\star}_a}\).}
\footnotetext[2]{More details can be found in our Appendix Section \ref{app:opti:design} and the work of \textcite{Prus2024} regarding \(Eff_{a}\). Note that \(\overline{Eff}_{a}\) refers to the mean of the 100 values involved.}
\footnotetext[3]{More details can be found in our Appendix Section \ref{app:opti:design} and the work of \textcite{Prus2024} regarding \(MSE_{\text{Tr}}\). Note that here \(\overline{MSE}_{\text{Tr}}\) refers to the mean of the 100  values involved.}

\clearpage

\section{Discussion}
Achieving proper convergence in the MCMC sampling algorithm to the posterior stationary distribution is essential for ensuring reliable posterior inference (\cite{Gelman2013}). Our posterior sampling algorithm faces certain convergence challenges, mainly occurring during the first window, while the subsequent windows tend to stabilize. However, as observed during the exploration of the algorithm, increasing the number of iterations during sampling generally improves convergence. Second, providing more data during the first window aids in achieving convergence. \\
We observed that sparse data in combination with uninformative priors can exacerbate convergence problems. Less informative priors require more data and a greater number of iterations to achieve convergence, while more informative priors can reduce this effort but carry the risk of bias if their information content is not justified by historical data. This points to a general issue in Bayesian workflows, as summarized in Table \ref{tab:informativeness}. A balance between computational feasibility and the informativeness of priors, especially when computational resources are limited, is essential. \\
Recall that our frequentist LMM can be seen as a specific case of the Bayesian LMM with uninformative uniform priors on the variance components (\cite{LairdWare1982}). If such a LMM with overly rudimentary prior assumptions is calculated, it can easily encounter unstable convergence, especially with high model complexity. Our hierarchical structured Bayesian LMM reformulation uses MCMC methods instead of REML estimation. With REML, variance component estimates often converge to exactly zero (\cite{Studnicki2024}). Estimates tend to become unstable near the boundary of the parameter space. We observed such variances for the effects var\_zone\_year and var\_gen\_zone\_year of our frequentist LMM specified in Equation \eqref{MM:specification:formular}. However, our MCMC approach stabilized convergence with the pragmatic solution to include a certain amount of prior information in a fully Bayesian framework utilizing historical data windows. This maintains consistent model accuracy, as all parameters assumed to be non-zero under the model specification are estimated properly. \\

\begin{table}[ht]
    \footnotesize
    \centering
    \begin{tabular}{llll}
            \toprule
            \textbf{Prior Type}               & \textbf{Convergence Risk} & \textbf{Stability} & \textbf{Bias Risk} \\
            \midrule
            Strongly Informative  & Low            & High & High \\
                                 &                &           & (can over-shrink estimates) \\
            Weakly Informative    & Medium            & Medium      & Low \\
                                 &                &           & (balances data and prior influence) \\
            Uninformative  & High & Low & None \\
            (flat, improper)      &                &      & (trends to divergence) \\
            \bottomrule
    \end{tabular}
    \caption{Comparison of prior types and their effects on convergence, stability, and bias risk inspired by \textcite{GelmanYao2020}.}
    \label{tab:informativeness}
\end{table}

Comparing the REML estimators from our frequentist LMM analysis with its MCMC based equivalent in Figures \ref{fig:post:histo:AllOther:cycle5:corr09} and \ref{fig:post:histo:GenZone:cycle5:corr09}, it appears that REML point estimates are not consistently embedded within regions of high posterior density. This underlines the instability issues of REML estimates. The Bayesian updating procedure provides substantially richer information in terms of uncertainty quantification, as well as positive and non-zero variance components in line with the underlying LMM assumption about them. Inverse gamma and Wishart priors deliver a clear formal starting point for demonstration purposes of our method. However, results for var\_year, var\_zone\_year, var\_zone\_loc\_year and var\_gen\_zone\_year in Figure \ref{fig:post:histo:AllOther:cycle5:corr09} hint at inverse gamma limitations, capturing high posterior density very close to zero. This becomes even more impressive with a closer look at Figure \ref{fig:post:histo:ALL:WINDOW:corr09}. Here, the REML estimate of the var\_gen\_year component is relatively small at 0.0246 and the REML estimate of the var\_zone\_year component is exactly zero. Over the windows, high posterior density regions shift more and more towards the REML point estimate, though they are not able to fully capture it. \\
Increasing the amount of data and the number of data windows shifts the initial starting point a priori more and more towards the observed MET-dataset. The whole MET dataset of interest should be divided, at least, into three historical data windows, each containing at least three years. However, the selected number of historical data windows should be carefully balanced according to the available data, particularly the number of years, the richness of genotype replication, and the number of environments evaluated per year. Although we selected hyper-parameters manually in this study, the framework performs well; future work may focus on developing a fully automated pipeline. \\
Our study highlights both the potential and the methodological challenges of Bayesian LMM frameworks for MET data, particularly when variance components are iteratively updated across historical windows. \textcite{przystalski2020, przystalski2023} investigated the prior specification of variance components, emphasizing the importance of parameter reparameterization. They recommend placing priors directly on standard deviations rather than on variance components. Following \textcite{Gelman2006}, weakly informative priors on standard deviations typically lead to more stable posterior samples and better MCMC convergence, particularly in hierarchical models with multiple random-effect layers. \\
Beyond reparameterization, several alternative prior strategies merit closer exploration. Joint multivariate normal–inverse gamma or normal–inverse Wishart priors for random effects and their covariance structures provide a promising alternative model structure that could lead to more efficient sampling. More flexible scalar priors such as half-$t$, half-Cauchy, improper uniform, or log-normal distributions for variance components (\cite{Gelman2006, Studnicki2024}) offer alternatives that can reduce bias in early updating windows. For correlation matrices, Lewandowski-Kurowicka-Joe priors (\cite{Lewandowski2009}) or scaled inverse Wishart priors (\cite{Alvarez2014}) may be preferable to the inverse Wishart due to their property of de-correlating the sampled chains of covariance matrix elements. \\
Further promising methodological extensions arise from exploring alternative variance-covariance structural assumptions of random effects. We recommend investigating factor-analytic (FA) structures as an alternative to US structures, which are widely used in advanced Bayesian MET modeling already (\cite{nuvunga2019}), to be reformulated in historical windows. Incorporating FA-based random-effect structures within the updating algorithm could substantially reduce dimensionality while preserving interpretable genetic covariance patterns. Alternatively, beta-mixed models (\cite{figueroa2013}) become relevant when modeling scaled or bounded MET traits, especially for stability indices or relative performance measures. \\
Efficient posterior approximation techniques such as the integrated nested Laplace approximation (INLA) (\cite{rue2009approximate, holand2013animal, mathew2015integrated}) or variational inference (VI) (\cite{yan2023}) represent promising alternatives to MCMC sampling in the presence of high spatial variances in individual field trials or very large MET datasets. \\

A major future direction concerns broadening the scope of applications for our Bayesian updating framework. In Table \ref{tab:bayes:design:with:year}, we demonstrated the usage of our method for obtaining A-optimal designs by \textcite{Prus2024}. Our findings align with those in their Tables 3 and 4. As J increases, the designs shift toward more balanced allocations, with balanced designs becoming increasingly efficient. However, in comparison to the frequentist approach of \textcite{Prus2024}, full uncertainty is taken into account in the average optimal allocations of trials \(\boldsymbol{\bar{\xi}^{\star}_a}\) in the approximate design. Consequently, \(\boldsymbol{\bar{\xi}^{\star}_a}\) tends to be more balanced. This is because extreme weight components in the posterior design sample tend to cancel each other out on average. Intuitively, this leads to higher efficiency values \(Eff_{a}\) and lower MSE matrix traces \(MSE_{\text{Tr}}\). \\
Moreover, our Bayesian updating of LMM variance parameters can be combined with spatial autoregressive models (\cite{gilmour1997}) or P-spline approaches (\cite{Rod.18, Piepho.21}) for single-trial spatial adjustment; extended stochastic simulation approaches, such as those carried out by \textcite{buntaran2022assessing} for responses to selection; and the integration of environmental covariates for genotype-specific predictions in new environments following \textcite{PiephoBlancon2023, tolhurst2022genomic}. In analogy to the information-loss issues observed in multi-stage analysis under frequentist approaches (\cite{Damesa2017}), Bayesian updating offers an intuitive mechanism to propagate uncertainty and dependence across stages, avoiding the collapse of variance components observed with point REML estimates. This ultimately leads to more reliable inference on variance component estimates. \\
While REML-based LMM remain the standard for MET analysis, their limitations in handling weak signals and complex random structures with large MET data motivate our provided Bayesian alternative. By leveraging historical data through Bayesian updating and connecting it to posterior uncertainty in optimal design, this study aims to advance the methodological toolkit available for MET. We would like to emphasize its relevance for both (i) breeding programs and (ii) variety testing, two closely related application areas of MET analysis, although our application with the data described in \textcite{Rahman2023} clearly refers to (ii). Companies conducting MET typically possess extensive historical data that can, in principle, improve variance component estimation and contribute to more reliable and informative plant breeding decisions. However, these data are rarely incorporated sufficiently into current year genomic selection processes. Our method avoids the need to re-estimate variance components of the entire historical MET data and enables convenient current-year updates.

\section{Acknowledgments}
We are grateful to the Bangladesh Rice Research Institute (BRRI) for generating, maintaining, and generously providing access to the long-term multi-environment trial data used in this study. The exploration of our statistical method would not have been possible without all the efforts of BRRI scientists, field staff, and technical personnel! \\
We thank Reinhard Meister for suggesting a fully Bayesian analysis of historical MET data at a seminar held by Hans-Peter Piepho at the Berliner Kolloquium “Statistische Methoden in der empirischen Forschung” in 1998. It has taken quite some time to get there, but better late than never!

\newpage

\printbibliography

\newpage

\section{Appendix}\label{supp:mat}
We provided a Git webpage at \href{https://github.com/StephanBark/Bayesian_Updating_with_Optimal_Design_2026}{GitHub}\footnote{\url{https://github.com/StephanBark/Bayesian_Updating_with_Optimal_Design_2026}} containing the BRRI MET data considered, our R and Stan code for the sampling algorithm, and the possibility to reproduce and extend the given results and model diagnostic plots. Note that we anonymized variety names from BRRI.

\subsection{Bayesian linear mixed model reformulation}
\label{app:bayesmod:bayes:cond}
Let \(i = 1, \ldots, n\) index the number of observations, and \(j = 1, \ldots, p\) index the number of fixed effects. Furthermore, let \(k = 1, \ldots, K+1\) denote the number of random effects, with \(q_k = 1, \ldots, Q_k\) indexing the levels of the random effect \(k\). Environments, defined as year-by-location combinations, are indexed by \(e = 1, \ldots, E\). For each environment, the total number of observations are \(N_e\). \\
Let \(\mathbf{y} \in \mathbb{R}^{n}\) denote the response vector, and \(\mathbf{X} \in \mathbb{R}^{n \times p}\) the fixed effect design matrix, with \(\boldsymbol{\beta} \in \mathbb{R}^p\) representing the vector of fixed effects. For each random effect \(k\), the design matrix is given by \(\mathbf{Z}_k \in \mathbb{R}^{n \times q_k}\), and the corresponding random effect vector is assumed to follow a multivariate normal distribution \(\mathbf{b}_k \sim \mathcal{N}(\mathbf{0}, \mathbf{G}_k)\), where \(\mathbf{G}_k\) is the covariance matrix of the \(k\)-th random effect. \\
For the genotype-zone random effect, the covariance structure is defined as \(\mathbf{G}_{K+1} = \mathbf{I}_{M} \otimes \mathbf{\Sigma}_{Z}\). This results in \(M\) identical \(4 \times 4\) covariance matrices \(\mathbf{\Sigma}_{Z}\) along the diagonal, each corresponding to one genotype, reflecting the US structural assumption of the zones in the genotype-zone random effect in the current dataset. For all other random effects, the covariance matrices are assumed to be of the form \(\mathbf{G}_k = \sigma^2_k \mathbf{I}_{q_k}\), where \(\sigma^2_k\) denotes the variance component associated with the random effect. The overall covariance matrix of the random effects is then given by the direct sum \(\mathbf{G} = \bigoplus_{k=1}^{K+1}\mathbf{G}_k\), which is the block-diagonal matrix with blocks \(\mathbf{G}_1, ..., \mathbf{G}_{K+1}\). \\
The residuals are modeled as \(\mathbf{\epsilon} \sim \mathcal{N}(\mathbf{0}, \mathbf{R})\), where \(\mathbf{R} = \bigoplus_{e=1}^{E}\mathbf{R}_e\) represents the heterogeneous residual covariance structure across environments. Each environment-specific covariance matrix is defined as \(\mathbf{R}_e = \sigma^2_e \mathbf{I}_{N_e}\), corresponding to environment \(e\). Finally, the vector of all random effects is denoted by \(\mathbf{b} = (\mathbf{b}_1, \ldots, \mathbf{b}_{K+1})^\top\).

\subsubsection{Linear mixed model in matrix notation}
\label{app:bayesmod:mixed:cond}
The model of the previous Section can now be expressed as
\begin{equation}
    \mathbf{y} \mid \mathbf{X}, \boldsymbol{\beta}, \mathbf{b}, \mathbf{R} \sim \mathcal{N} \left( \mathbf{X} \boldsymbol{\beta} + \sum_{k=1}^{K+1} \mathbf{Z}_k \mathbf{b}_k, \mathbf{R} \right),
\end{equation}
where \(\text{Cov}(\mathbf{y} \mid \mathbf{b}) = \mathbf{R}\) reflects the covariance structure of the response. \\
The corresponding likelihood function can be formulated as
\begin{equation}
    p(\mathbf{y} \mid \mathbf{X}, \boldsymbol{\beta}, \mathbf{G}, \mathbf{R}) = \frac{1}{\sqrt{(2\pi)^n \det\left( \mathbf{R} \right)}} \times \nonumber
\end{equation}
\begin{equation}
    \exp \left( -\frac{1}{2} (\mathbf{y} - \mathbf{X} \boldsymbol{\beta} - \sum_{k=1}^{K+1} \mathbf{Z}_k \mathbf{b}_k)^\top \mathbf{R}^{-1} (\mathbf{y} - \mathbf{X} \boldsymbol{\beta} - \sum_{k=1}^{K+1} \mathbf{Z}_k \mathbf{b}_k) \right).
\end{equation} 

\subsubsection{Prior assumptions specification}
\label{app:cond:MM:prior}
For the fixed effects vector \(\boldsymbol{\beta}\), by default, \texttt{rstan} package assumes that \(\boldsymbol{\beta} \sim p(\boldsymbol{\beta})\), where \(p(\boldsymbol{\beta})\) corresponds to an improper uninformative uniform prior that supports the entire real line. For the random effects corresponding to the genotype-zone random effects, it is assumed that \(\mathbf{b}_{K+1} \sim p(\mathbf{b}_{K+1})\), where \(p(\mathbf{b}_{K+1})\) is a multivariate normal distribution \(\mathcal{N}(\mathbf{0}, \mathbf{G}_{K+1})\). Recall that \(\mathbf{G}_{K+1}\) has block diagonal structure, reflecting the US covariance assumption across zones for each genotype. For all other random effects, the assumption is that \(\mathbf{b}_{k} \sim p(\mathbf{b}_{k})\), where \(p(\mathbf{b}_{k})\) is a multivariate normal distribution \(\mathcal{N}(\mathbf{0}, \mathbf{G}_{k})\). In this case, each \(\mathbf{G}_{k}\) is known to have a diagonal structure that corresponds to independent variance components. For the covariance matrix corresponding to the zones of the genotype-zone random effect, it is assumed that \(\mathbf{\Sigma}_{Z} \sim p(\mathbf{\Sigma}_{Z})\), where \(p(\mathbf{\Sigma}_{Z})\) is an inverse Wishart distribution \(\text{Inv-Wishart}(\nu, \mathbf{S})\) with \(\nu > Z-1\) degrees of freedom and a positive definite scale matrix \(\mathbf{S}\). For all other random effect variance parameters, it is assumed that \(\sigma^2_k \sim p(\sigma^2_k)\), where each \(p(\sigma^2_k)\) is independently an inverse gamma distribution \(\text{Inv-Gamma}(\alpha_k, \beta_k)\) with the shape parameter \(\alpha_k > 0\) and the scale parameter \(\beta_k > 0\). Finally, for all environment-specific residual variance components, it is assumed that \(\sigma^2_e \sim p(\sigma^2_e)\), where each \(p(\sigma^2_e)\) is independently an inverse gamma distribution \(\text{Inv-Gamma}(\alpha_e, \beta_e)\) with the shape parameter \(\alpha_e > 0\) and the scale parameter \(\beta_e > 0\). These prior assumptions provide the necessary distributional framework for Bayesian inference of fixed effects, random effects, and variance components in the proposed model.

\subsubsection{Joint posterior formulation}
\label{app:cond:MM:joint}
The joint-posterior distribution can be described as
\begin{equation}
    p(\boldsymbol{\beta}, \mathbf{b}, \boldsymbol{\sigma}^2_{k}, \mathbf{\Sigma}_Z, \boldsymbol{\sigma}^2_e \mid  \mathbf{X}, \mathbf{y}) \propto \nonumber
\end{equation}
\begin{equation}
    p(\mathbf{y} \mid \mathbf{X}, \boldsymbol{\beta}, \mathbf{b}, \mathbf{G}, \mathbf{R}) \times p(\boldsymbol{\beta}) \times \prod_{k=1}^{K+1} p(\mathbf{b}_k \mid \mathbf{G}_k) \times \prod_{k=1}^K p(\sigma^2_k) \times p(\mathbf{\Sigma}_{Z}) \times \prod_{e=1}^{E} p(\sigma^2_{e}) \propto \nonumber
\end{equation}
\begin{equation}
    \frac{1}{\sqrt{\det\left( \mathbf{R} \right)}} \exp \left( -\frac{1}{2} (\mathbf{y} - \mathbf{X} \boldsymbol{\beta} - \sum_{k=1}^{K+1} \mathbf{Z}_k \mathbf{b}_k)^\top \mathbf{R}^{-1} (\mathbf{y} - \mathbf{X} \boldsymbol{\beta} - \sum_{k=1}^{K+1} \mathbf{Z}_k \mathbf{b}_k) \right)
    \times \nonumber
\end{equation}
\begin{equation}
    \prod_{k=1}^{K+1} \frac{1}{\sqrt{\det\left( \mathbf{G}_k \right)}} \exp \left( -\frac{1}{2} \mathbf{b}_k^\top \mathbf{G}_k^{-1} \mathbf{b}_k \right)
    \times \nonumber
\end{equation}
\begin{equation}
    \prod_{k=1}^{K} \frac{\beta_k^{\alpha_k}}{\Gamma(\alpha_k)} (\sigma_k^2)^{-\alpha_k-1} \exp\left(-\frac{\beta_k}{\sigma_k^2}\right)
    \times \nonumber
\end{equation}
\begin{equation}
    \frac{\det(\mathbf{S})^{\nu/2}}{2^{(\nu Z)/2} \Gamma_Z\left(\frac{\nu}{2}\right)} (\det(\mathbf{\Sigma}_{Z}))^{-(\nu + Z + 1)/2} \exp\left( -\frac{1}{2} \text{tr}(\mathbf{S} \mathbf{\Sigma}_{Z}^{-1}) \right) \times \nonumber
\end{equation}
\begin{equation}
    \prod_{e=1}^{E} \frac{\beta_e^{\alpha_e}}{\Gamma(\alpha_e)} (\sigma_e^2)^{-\alpha_e-1} \exp\left(-\frac{\beta_e}{\sigma_e^2}\right),
\end{equation}
where \(\boldsymbol{\sigma}^2_{k} = (\sigma^2_{1}, \dots, \sigma^2_{K})^\top\) and \(\boldsymbol{\sigma}^2_e = (\sigma^2_{1}, \dots, \sigma^2_{E})^\top\) and \(\Gamma(\cdot)\) is the Gamma function for the Inverse-Gamma distribution and \(\Gamma_Z(\cdot)\) is the multivariate Gamma function for the inverse Wishart distribution and \(\text{tr}(\cdot)\) denotes the matrix trace operator.

\subsubsection{Full conditional posterior specifications}
\label{app:bayesmod:mixed:fullcond2}
The full conditional posterior of any \(\mathbf{b}_{\tilde{k}}\) can be expressed as
\begin{equation}
    p(\mathbf{b}_{\tilde{k}} \mid \boldsymbol{\beta}, \mathbf{b}_{-{\tilde{k}}}, \mathbf{G}, \mathbf{R},  \mathbf{X}, \mathbf{y}) \stackrel{Bayes}{=} \nonumber
\end{equation}
\begin{equation}
    \frac{p(\boldsymbol{\beta}, \mathbf{b}, \boldsymbol{\sigma}^2_{k}, \mathbf{\Sigma}_Z, \boldsymbol{\sigma}^2_e \mid  \mathbf{X}, \mathbf{y})}{p(\boldsymbol{\beta}, \mathbf{b}_{-\tilde{k}}, \boldsymbol{\sigma}^2_{k}, \mathbf{\Sigma}_Z, \boldsymbol{\sigma}^2_e \mid  \mathbf{X}, \mathbf{y})} \stackrel{wrt. \mathbf{b}_{\tilde{k}}}{\propto} p(\boldsymbol{\beta}, \mathbf{b}, \boldsymbol{\sigma}^2_{k}, \mathbf{\Sigma}_Z, \boldsymbol{\sigma}^2_e \mid  \mathbf{X}, \mathbf{y}) \propto \nonumber
\end{equation}
\begin{equation}
    \frac{1}{\sqrt{\det\left( \mathbf{R} \right)}} \exp \left( -\frac{1}{2} (\mathbf{y} - \mathbf{X} \boldsymbol{\beta} - \sum_{k=1}^{K+1} \mathbf{Z}_k \mathbf{b}_k)^\top \mathbf{R}^{-1} (\mathbf{y} - \mathbf{X} \boldsymbol{\beta} - \sum_{k=1}^{K+1} \mathbf{Z}_k \mathbf{b}_k) \right)
    \times \nonumber
\end{equation}
\begin{equation}
    \prod_{k=1}^{K+1} \frac{1}{\sqrt{\det\left( \mathbf{G}_k \right)}} \exp \left( -\frac{1}{2} \mathbf{b}_k^\top \mathbf{G}_k^{-1} \mathbf{b}_k \right)
    \propto \nonumber
\end{equation}
\begin{equation}
    \exp \left( -\frac{1}{2} (\mathbf{y}_{\tilde{k}} - \mathbf{Z}_{\tilde{k}} \mathbf{b}_{\tilde{k}})^\top \mathbf{R}^{-1} (\mathbf{y}_{\tilde{k}} - \mathbf{Z}_{\tilde{k}} \mathbf{b}_{\tilde{k}}) \right)
    \times \exp \left( -\frac{1}{2} \mathbf{b}_{\tilde{k}}^\top \mathbf{G}_{\tilde{k}}^{-1} \mathbf{b}_{\tilde{k}} \right)
    \propto \nonumber
\end{equation}
\begin{equation}
    \exp \left( -\frac{1}{2} \left[ -2\mathbf{y}_{\tilde{k}}^\top \mathbf{R}^{-1} \mathbf{Z}_{\tilde{k}} \mathbf{b}_{\tilde{k}} + (\mathbf{Z}_{\tilde{k}} \mathbf{b}_{\tilde{k}})^\top \mathbf{R}^{-1} \mathbf{Z}_{\tilde{k}} \mathbf{b}_{\tilde{k}} + \mathbf{b}_{\tilde{k}}^\top \mathbf{G}_{\tilde{k}}^{-1} \mathbf{b}_{\tilde{k}} \right] \right) \propto \nonumber
\end{equation}
\begin{equation}
    \exp \left( -\frac{1}{2} \left[ \mathbf{b}_{\tilde{k}}^\top \underbrace{\left( \mathbf{Z}_{\tilde{k}}^\top \mathbf{R}^{-1} \mathbf{Z}_{\tilde{k}} + \mathbf{G}_{\tilde{k}}^{-1}\right)}_{\mathbf{\Omega}_{\tilde{k}}^{-1}} \mathbf{b}_{\tilde{k}} - 2 \mathbf{b}_{\tilde{k}}^\top \mathbf{\Omega}_{\tilde{k}}^{-1} \underbrace{\mathbf{\Omega}_{\tilde{k}} \mathbf{Z}_{\tilde{k}}^\top \mathbf{R}^{-1} \mathbf{y}_{\tilde{k}}}_{\mathbf{\tilde{b}}_{\tilde{k}}} \right] \right) \propto \nonumber
\end{equation}
\begin{equation}
    \exp \left( -\frac{1}{2} \left[ \mathbf{b}_{\tilde{k}}^\top \mathbf{\Omega}_{\tilde{k}}^{-1} \mathbf{b}_{\tilde{k}} - 2 \mathbf{b}_{\tilde{k}}^\top \mathbf{\Omega}_{\tilde{k}}^{-1} \mathbf{\tilde{b}}_{\tilde{k}} + \mathbf{\tilde{b}}_{\tilde{k}}^\top \mathbf{\Omega}_{\tilde{k}}^{-1} \mathbf{\tilde{b}}_{\tilde{k}} \right] \right) = \nonumber
\end{equation}
\begin{equation}
    \exp \left( -\frac{1}{2} \left[ (\mathbf{b}_{\tilde{k}} - \mathbf{\tilde{b}}_{\tilde{k}})^\top \mathbf{\Omega}_{\tilde{k}}^{-1} (\mathbf{b}_{\tilde{k}} - \mathbf{\tilde{b}}_{\tilde{k}}) \right] \right),
\end{equation}
where \(\mathbf{b}_{-\tilde{k}} = \bigoplus_{k \neq \tilde{k}}\mathbf{b}_k\), which is a vector that consists of components of all vectors \(b_1, .., b_{K+1}\) except of \(b_{\tilde{k}}\) and \(\mathbf{y}_{\tilde{k}} = \mathbf{y} - \mathbf{X} \boldsymbol{\beta} - \sum_{k \neq \tilde{k}} \mathbf{Z}_k \mathbf{b}_k\). Note that the aim of the above calculation is to derive the kernel of the full conditional posterior of \(\boldsymbol{b}_{\tilde{k}}\). In that context, every factor that is constant in the \(\boldsymbol{b}_{\tilde{k}}\) parameter vector of interest can be dropped up to proportionality. E.g., the factor \(\exp \left( -\frac{1}{2} \mathbf{y}_{\tilde{k}}^\top \mathbf{R}^{-1} \mathbf{y}_{\tilde{k}} \right)\) can be dropped for that purpose. Furthermore, with respect to proportionality, the normalizing constant of the full conditional posterior kernel of \(\boldsymbol{b}_{\tilde{k}}\) can be multiplied on top after the posterior parameters are derived. This corresponds to a kernel of a normal distribution: 
\begin{equation}
    \mathbf{b}_{\tilde{k}} \mid \boldsymbol{\beta}, \mathbf{b}_{-{\tilde{k}}}, \mathbf{G}, \mathbf{R}, \mathbf{y} \sim \mathcal{N} \left( \mathbf{\tilde{b}}_{\tilde{k}},\mathbf{\Omega}_{\tilde{k}} \right).
\end{equation}
The full conditional posterior for any \(\sigma^2_{\tilde{k}}\) for which \(\mathbf{G}_{\tilde{k}} = \sigma^2_{\tilde{k}} \mathbf{I}_{q_{\tilde{k}}}\) is given below.
\begin{equation}
    p(\sigma^2_{\tilde{k}} \mid \boldsymbol{\beta}, \mathbf{b}_k, \mathbf{G}_{-\tilde{k}}, \mathbf{R},  \mathbf{X}, \mathbf{y}) \stackrel{Bayes}{=} \nonumber
\end{equation}
\begin{equation}
    \frac{p(\boldsymbol{\beta}, \mathbf{b}_k, \boldsymbol{\sigma}^2_{k}, \mathbf{\Sigma}_Z, \boldsymbol{\sigma}^2_e \mid  \mathbf{X}, \mathbf{y})}{p(\boldsymbol{\beta}, \mathbf{b}_k, \boldsymbol{\sigma}^2_{-\tilde{k}}, \mathbf{\Sigma}_Z, \boldsymbol{\sigma}^2_e \mid  \mathbf{X}, \mathbf{y})} \stackrel{wrt. \sigma^2_{\tilde{k}}}{\propto} p(\boldsymbol{\beta}, \mathbf{b}_k, \boldsymbol{\sigma}^2_{k}, \mathbf{\Sigma}_Z, \boldsymbol{\sigma}^2_e \mid  \mathbf{X}, \mathbf{y}) \propto  \nonumber
\end{equation}
\begin{equation}
    \frac{1}{\sqrt{\det\left( \sigma^2_{\tilde{k}} \mathbf{I}_{Q_{\tilde{k}}} \right)}} \times \exp \left( -\frac{1}{2} \mathbf{b}_{\tilde{k}}^\top \left( \sigma^2_{\tilde{k}} \mathbf{I}_{Q_{\tilde{k}}} \right)^{-1} \mathbf{b}_{\tilde{k}} \right) (\sigma_{\tilde{k}}^2)^{-\alpha_{\tilde{k}}-1} \exp\left(-\frac{\beta_{\tilde{k}}}{\sigma_{\tilde{k}}^2}\right) \propto \nonumber
\end{equation}
\begin{equation}
    (\sigma_{\tilde{k}}^2)^{-(\alpha_{\tilde{k}} + \frac{Q_{\tilde{k}}}{2}) - 1} \times \exp \left( - \frac{1}{\sigma^2_{\tilde{k}}} \left[ \beta_{\tilde{k}} + \frac{1}{2} \mathbf{b}_{\tilde{k}}^\top \mathbf{b}_{\tilde{k}} \right]\right),
\end{equation}
where \(\mathbf{G}_{-\tilde{k}} = \bigoplus_{k\neq \tilde{k}}^{K+1}\mathbf{G}_k \) and \(\boldsymbol{\sigma}^2_{-\tilde{k}} = (\sigma^2_1, \dots, \sigma^2_{\tilde{k}-1}, \sigma^2_{\tilde{k}+1}, \dots, \sigma^2_K)^\top\). \\

This corresponds to a kernel of an inverse gamma distribution:
\begin{equation}
    \sigma_{\tilde{k}}^2 \mid \boldsymbol{\beta}, \mathbf{b}_{k}, \mathbf{G}_{-\tilde{k}}, \mathbf{R}, \mathbf{y} \sim \text{Inv-Gamma}( \tilde{\alpha}_{\tilde{k}}, \tilde{\beta}_{\tilde{k}}),
\end{equation}
where \(\tilde{\alpha}_{\tilde{k}} = \alpha_{\tilde{k}} + \frac{Q_{\tilde{k}}}{2}\) and \(\tilde{\beta}_{\tilde{k}} = \beta_{\tilde{k}} + \frac{1}{2} \mathbf{b}_{\tilde{k}}^\top \mathbf{b}_{\tilde{k}}\). \\

The full conditional posterior for \(\mathbf{\Sigma}_Z\) for which \(\mathbf{G}_{\tilde{k}} = \mathbf{I}_{M} \otimes \mathbf{\Sigma}_{Z}\) is given below.
\begin{equation}
    p(\mathbf{\Sigma}_Z \mid \boldsymbol{\beta}, \mathbf{b}_k, \mathbf{G}_{-\tilde{k}}, \mathbf{R},  \mathbf{X}, \mathbf{y}) \stackrel{Bayes}{=} \nonumber
\end{equation}
\begin{equation}
    \frac{p(\boldsymbol{\beta}, \boldsymbol{\sigma}^2_{k}, \mathbf{\Sigma}_Z, \boldsymbol{\sigma}^2_e \mid \mathbf{b}_k,  \mathbf{X}, \mathbf{y})}{p(\boldsymbol{\beta}, \boldsymbol{\sigma}^2_{k}, \boldsymbol{\sigma}^2_e \mid \mathbf{b}_k,  \mathbf{X}, \mathbf{y})} \stackrel{wrt. \mathbf{\Sigma}_Z}{\propto} p(\boldsymbol{\beta}, \boldsymbol{\sigma}^2_{k}, \mathbf{\Sigma}_Z, \boldsymbol{\sigma}^2_e \mid \mathbf{b}_k,  \mathbf{X}, \mathbf{y}) \propto \nonumber
\end{equation}
\begin{equation}
    \frac{1}{\sqrt{\det\left( \mathbf{I}_{M} \otimes \mathbf{\Sigma}_{Z} \right)}} \times \exp \left( -\frac{1}{2} \mathbf{b}_{\tilde{k}}^\top \left(\mathbf{I}_{M} \otimes \mathbf{\Sigma}_{Z} \right)^{-1} \mathbf{b}_{\tilde{k}} \right) \times \nonumber
\end{equation}
\begin{equation} 
    \det(\mathbf{\Sigma}_{Z})^{-(\nu + Z + 1)/2} \exp\left( -\frac{1}{2} \text{tr}(\mathbf{S} \mathbf{\Sigma}_{Z}^{-1}) \right) \propto \nonumber
\end{equation}
\begin{equation}
    \det(\mathbf{\Sigma}_{Z})^{-(\nu + Z + M + 1)/2} \exp \left( -\frac{1}{2} \left[\text{tr}(\mathbf{S} \mathbf{\Sigma}_{Z}^{-1}) +  \mathbf{b}_{\tilde{k}}^\top \left( \mathbf{I}_{M} \otimes \mathbf{\Sigma}_{Z}^{-1} \right) \mathbf{b}_{\tilde{k}} \right] \right) = \nonumber
\end{equation}
\begin{equation}
    \det(\mathbf{\Sigma}_{Z})^{-(\nu + Z + M + 1)/2} \exp \left( -\frac{1}{2} \left[\text{tr}(\mathbf{S} \mathbf{\Sigma}_{Z}^{-1}) + \sum_{g=1}^{M} \mathbf{b}_{\tilde{k};g}^\top \mathbf{\Sigma}_{Z}^{-1} \mathbf{b}_{\tilde{k};g} \right] \right) = \nonumber
\end{equation}
\begin{equation}
    \det(\mathbf{\Sigma}_{Z})^{-(\nu + Z + M + 1)/2} \exp \left( -\frac{1}{2} \left[\text{tr}(\mathbf{S} \mathbf{\Sigma}_{Z}^{-1}) + \text{tr} \left( \mathbf{\Sigma}_{Z}^{-1} \sum_{g=1}^{M} \mathbf{b}_{\tilde{k};g}  \mathbf{b}_{\tilde{k};g}^\top \right) \right] \right) = \nonumber
\end{equation}
\begin{equation} 
    \det(\mathbf{\Sigma}_{Z})^{-(\nu + Z + M + 1)/2} \exp \left( -\frac{1}{2} \left(\text{tr}\left[\mathbf{\Sigma}_{Z}^{-1}\left(\mathbf{S} + \sum_{g=1}^{M} \mathbf{b}_{\tilde{k};g}  \mathbf{b}_{\tilde{k};g}^\top \right) \right] \right) \right),
\end{equation}
where \(\mathbf{b}_{\tilde{k}} = (\mathbf{b}_{\tilde{k};1}^\top, \dots, \mathbf{b}_{\tilde{k};M}^\top)^\top\). This corresponds to a kernel of an inverse Wishart distribution:
\begin{equation}
    \mathbf{\Sigma}_{Z} \mid \boldsymbol{\beta}, \mathbf{b}_k, \mathbf{G}_{-\tilde{k}}, \mathbf{R}, \mathbf{y} \sim \text{Inv-Wishart}(\tilde{\nu}, \tilde{\mathbf{S}}),
\end{equation} 
where \(\tilde{\nu} = \nu + M\) and \(\tilde{\mathbf{S}} = \mathbf{S} + \sum_{g = 1}^{M} \mathbf{b}_{\tilde{k};g}\mathbf{b}_{\tilde{k};g}^\top.\) \\

The full conditional posterior for any \(\sigma^2_{\tilde{e}}\) is given below.
\begin{equation}
    p(\sigma^2_{\tilde{e}} \mid \boldsymbol{\beta}, \mathbf{b}_k, \mathbf{G}, \mathbf{R}_{-\tilde{e}},  \mathbf{X}, \mathbf{y}) \stackrel{Bayes}{=} \nonumber
\end{equation}
\begin{equation}
    \frac{p(\boldsymbol{\beta}, \mathbf{b}_k, \boldsymbol{\sigma}^2_{k}, \mathbf{\Sigma}_Z, \boldsymbol{\sigma}^2_e \mid  \mathbf{X}, \mathbf{y})}{p(\boldsymbol{\beta}, \mathbf{b}_k, \boldsymbol{\sigma}^2_{k}, \mathbf{\Sigma}_Z, \boldsymbol{\sigma}^2_{-\tilde{e}} \mid  \mathbf{X}, \mathbf{y})} \stackrel{wrt. \sigma^2_{\tilde{e}}}{\propto} p(\boldsymbol{\beta}, \mathbf{b}_k, \boldsymbol{\sigma}^2_{k}, \mathbf{\Sigma}_Z, \boldsymbol{\sigma}^2_e \mid  \mathbf{X}, \mathbf{y}) \propto \nonumber
\end{equation}
\begin{equation}
    \frac{1}{\sqrt{\det\left( \bigoplus_{e=1}^{E}\sigma^2_e\mathbf{I}_{N_e} \right)}} \times \nonumber
\end{equation}
\begin{equation}
    \exp \left( -\frac{1}{2} (\mathbf{y} - \mathbf{X} \boldsymbol{\beta} - \sum_{k=1}^{K+1} \mathbf{Z}_k \mathbf{b}_k)^\top \left( \bigoplus_{e=1}^{E}\sigma^2_e\mathbf{I}_{N_e} \right)^{-1} (\mathbf{y} - \mathbf{X} \boldsymbol{\beta} - \sum_{k=1}^{K+1} \mathbf{Z}_k \mathbf{b}_k) \right) \times \nonumber
\end{equation}
\begin{equation}
    (\sigma_{\tilde{e}}^2)^{-\alpha_{\tilde{e}}-1} \exp\left(-\frac{\beta_{\tilde{e}}}{\sigma_{\tilde{e}}^2}\right) \propto \nonumber
\end{equation}
\begin{equation}
    (\sigma_{\tilde{e}}^2)^{-(\alpha_{\tilde{e}} + \frac{N_{\tilde{e}}}{2}) - 1} \times \nonumber 
\end{equation}
\begin{equation}
    \exp \left( -\frac{1}{2} \sum_{e=1}^E(\mathbf{y}_e - \mathbf{X}_e \boldsymbol{\beta} - \sum_{k=1}^{K+1} \mathbf{Z}_{k;e} \mathbf{b}_{k})^\top \left( \sigma^2_e\mathbf{I}_{n_{e}} \right)^{-1} (\mathbf{y}_e - \mathbf{X}_e \boldsymbol{\beta} - \sum_{k=1}^{K+1} \mathbf{Z}_{k;e} \mathbf{b}_{k}) -\frac{\beta_{\tilde{e}}}{\sigma_{\tilde{e}}^2}\right) \propto \nonumber
\end{equation}
\begin{equation}
    (\sigma_{\tilde{e}}^2)^{-(\alpha_{\tilde{e}} + \frac{N_{\tilde{e}}}{2}) - 1} \times \nonumber
\end{equation}
\begin{equation}
    \exp \left( -\frac{1}{\sigma^2_{\tilde{e}}} \left[\beta_{\tilde{e}} + \frac{1}{2}(\mathbf{y}_{\tilde{e}} - \mathbf{X}_{\tilde{e}} \boldsymbol{\beta} - \sum_{k=1}^{K+1} \mathbf{Z}_{k;{\tilde{e}}} \mathbf{b}_{k})^\top (\mathbf{y}_{\tilde{e}} - \mathbf{X}_{\tilde{e}} \boldsymbol{\beta} - \sum_{k=1}^{K+1} \mathbf{Z}_{k;{\tilde{e}}} \mathbf{b}_{k}) \right]\right).
\end{equation}
This corresponds to a kernel of an inverse gamma distribution:
\begin{equation}
    \sigma^2_{\tilde{e}} \mid \boldsymbol{\beta}, \mathbf{b}_k, \mathbf{G}, \mathbf{R}_{-{{\tilde{e}}}}, \mathbf{y} \sim \text{Inv-Gamma}(\tilde{\alpha}_{\tilde{e}}, \tilde{\beta}_{\tilde{e}}),
\end{equation}
with \(\tilde{\alpha}_{\tilde{e}} = \alpha_{\tilde{e}} + \frac{N_{\tilde{e}}}{2}\) and \(\tilde{\beta}_{\tilde{e}} = \beta_{\tilde{e}} + \frac{1}{2}(\mathbf{y}_{\tilde{e}} - \mathbf{X}_{\tilde{e}} \boldsymbol{\beta} - \sum_{k=1}^{K+1} \mathbf{Z}_{k;{\tilde{e}}} \mathbf{b}_{k})^\top (\mathbf{y}_{\tilde{e}} - \mathbf{X}_{\tilde{e}} \boldsymbol{\beta} - \sum_{k=1}^{K+1} \mathbf{Z}_{k;{\tilde{e}}} \mathbf{b}_{k})\). It can be concluded that for the BLMM based on a LMM perspective, all parameters of interest have known full conditional posterior distributions.

\subsection{Multi-year window Bayesian updating sampling algorithm}
\label{app:bayesmod:bayes:sampling}
To make the sampling algorithm proper, the structural assumptions in the prior distributions have to be achieved in the posterior distribution as well. While \(p(\boldsymbol{\beta})\), \(\prod_{e=1}^{E} p(\sigma^2_{e})\), \(\prod_{k=1}^K p(\sigma^2_k)\), and \(p(\mathbf{\Sigma}_{Z})\) are marginal prior distributions, \(\prod_{k=1}^{K+1} p(\mathbf{b}_k \mid \mathbf{G}_k)\) is a conditional prior distribution. The underlying dependency of random effects and their corresponding variance components has to be taken into account during the sampling process. It is desired to keep the LMM framework with its structural assumptions during the window informing process to be able to use the A-optimality design criterion of \textcite{Prus2024} later in the pipeline. Their criteria are especially derived from a specific LMM. \\ 
The 22-year dataset described in \textcite{Rahman2023} is split up into smaller historical datasets. Justified by proper convergence, we chose six years in the oldest historical dataset and three years for the following datasets, which leads to a total of five datasets.

\subsubsection{Maximum likelihood based posterior parameter estimates}
\label{app:cond:MCMC:MLE}
In order to estimate and carry the posterior parameters through every consecutive multi-year window, the maximum likelihood estimates (MLE) of the distributional parameters were calculated using the drawn posterior sample in each window. This approach is valid under certain approximation assumptions. The drawn marginal Gibbs sample of each variance parameter is assumed to be approximately from an inverse Gamma or inverse Wishart distribution. The validity of this approximation can be verified by the marginal posterior histogram plots reproducible in the Git webpage of this study. Note that in the window procedure described in this section, the posterior parameters and distributions of the random and fixed effects are not the primary focus. However, they play a crucial implicit role in informing the prior distributions of the variance components. Only the estimated posterior parameters of the variance component marginal distributions are carried through the windows. \\
Consider a posterior sample with sampled values \(t = 1, \dots, B, \dots, T\) where \(B\) corresponds to the number of burn-in draws. In the MCMC context, strictly speaking, the i.i.d. assumption usually assumed in ML calculations, as given below, is violated. However, the ergodic theorem justifies that the calculations are still proper when the samples are drawn - without considering burn-in draws - from the stationary distribution after convergence. Finally, note that the following optimization problems are in practice transferred to minimization problems of the sum of the negative log-likelihood of the sample components.
\begin{equation}
    \left[ (\widehat{\alpha}_{k})^{(\text{MLE})}, (\widehat{\beta}_{k})^{(\text{MLE})} \right]^{\top} = \underset{[\alpha_{k}, \beta_{k} > 0]}{\text{argmax}} \ \mathcal{L}\left( \alpha_{k}, \beta_{k} \mid  (\boldsymbol{\sigma}^2_{k})^{(\text{Post})} \right)
\end{equation}
\(\forall \ k = 1, \dots, K\), and with \((\boldsymbol{\sigma}^2_k)^{(\text{Post})} = \left[ (\sigma^2_k)^{(B+1)}, \dots, (\sigma^2_k)^{(T)} \right]^{\top}\) as the marginal posterior sample corresponding to \(\sigma^2_k\), and \(\mathcal{L}\left( \alpha_{k}, \beta_{k} \mid  (\boldsymbol{\sigma}^2_{k})^{(\text{Post})} \right)\) corresponds to an inverse gamma likelihood with the distributional parameters \(\alpha_k\) and \(\beta_k\), and with the data vector \((\boldsymbol{\sigma}^2_{k})^{(\text{Post})}\).
\begin{equation}
    \left[ (\widehat{\nu})^{(\text{MLE})}, (\widehat{\mathbf{S}})^{(\text{MLE})} \right]^{\top} = \underset{[\nu > Z+1 \ \text{and} \ \mathbf{S} \ \text{pos. def.}]}{\text{argmax}} \ \mathcal{L}\left( \nu, \mathbf{S} \mid (\mathbf{\Sigma}_{Z})^{(\text{Post})} \right)
\end{equation}
with \((\mathbf{\Sigma}_{Z})^{(\text{Post})} = \left[ (\mathbf{\Sigma}_{Z})^{(B+1)}, \dots, (\mathbf{\Sigma}_{Z})^{(T)} \right]^{\top}\) as the marginal posterior sample corresponding to \(\mathbf{\Sigma}_Z\), and \(\mathcal{L}\left( \nu, \mathbf{S} \mid (\mathbf{\Sigma}_{Z})^{(\text{Post})} \right)\) corresponds to an inverse Wishart likelihood with the distributional parameters \(\nu\) and \(\mathbf{S}\), and with the data vector \((\mathbf{\Sigma}_{Z})^{(\text{Post})}\).
\begin{equation}
    \left[ (\widehat{\alpha}_{e})^{(\text{MLE})}, (\widehat{\beta}_{e})^{(\text{MLE})} \right]^{\top} = \underset{[\alpha_{e}, \beta_{e} > 0]}{\text{argmax}} \ \mathcal{L}\left( \alpha_{e}, \beta_{e} \mid (\boldsymbol{\sigma}^2_e)^{(\text{Post})} \right)
\end{equation}
\(\forall \ e = 1, \dots, E\), and with \((\boldsymbol{\sigma}^2_e)^{(\text{Post})} = \left[ (\sigma^2_e)^{(B+1)}, \dots, (\sigma^2_e)^{(T)} \right]^{\top}\) as the marginal posterior sample corresponding to \(\sigma^2_e\), and \(\mathcal{L}\left( \alpha_{e}, \beta_{e} \mid (\boldsymbol{\sigma}^2_e)^{(\text{Post})} \right)\) corresponds to an inverse gamma likelihood with the distributional parameters \(\alpha_e\) and \(\beta_e\), and with the data vector \((\boldsymbol{\sigma}^2_e)^{(\text{Post})}\).

\subsubsection{Pseudo code of the Bayesian updating sampling algorithm}
\label{app:algo:pseudocode:main}
This approach uses the derivations of the BLMM Specification Section \ref{app:bayesmod:bayes:cond}. Here, all full conditional posterior distributions are known. In addition, it is assumed that the marginal posterior Gibbs samples of the variance components are approximately inverse Gamma or inverse Wishart distributions. To consider the conditioning in \(p(\mathbf{b} \mid \mathbf{G})\) on \(\mathbf{G}\) only the information in the posterior parameters of the marginal distribution of the variance components is passed to the next window. All variance component posterior parameters are estimated via ML estimation based on their drawn marginal posterior Gibbs samples as described in the previous section. In the following, the algorithm for the iterative multi-year window sampling procedure is described as pseudo code. For that, consider \(t = 1, \dots, B, \dots, T\) posterior sampled values in each multi-year window, where \(B\) corresponds to the number of burn-in draws. For each multi-year window dataset \(\mathbf{y}_l\) with \(l = 1, \dots, L\) it holds:
\begin{enumerate}
    \item[(1)] Initialize \( (\boldsymbol{\Psi}_1)^{(0)}_{(l)} = (\boldsymbol{\beta}^{(0)}_{(l)}, \mathbf{b}^{(0)}_{(l)}, (\boldsymbol{\sigma}^2_{k})^{(0)}_{(l)}, (\mathbf{\Sigma}_Z)^{(0)}_{(l)}, (\boldsymbol{\sigma}^2_e)^{(0)}_{(l)})^\top \) as draws from the corresponding priors.
    \item[(2)] For each posterior sample component \( t = 1, \dots, B, \dots, T \):
    \begin{itemize}
        \item[(2.1)] sample \( \boldsymbol{\beta}_{(l)}^{(t)} \sim p_{(l)}(\boldsymbol{\beta} | (\mathbf{b})_{(l)}^{(t-1)}, (\boldsymbol{\sigma}^2_{k})_{(l)}^{(t-1)}, (\mathbf{\Sigma}_Z)_{(l)}^{(t-1)}, (\boldsymbol{\sigma}^2_e)_{(l)}^{(t-1)}, \mathbf{X}_{(l)}, \mathbf{y}_{(l)}) \),
        \item[(2.2)] sample for \(k = 1, \dots, K+1\): \\
        \( (\mathbf{b}_{k})_{(l)}^{(t)} \sim \\
        p_{(l)}(\mathbf{b}_{k} | \boldsymbol{\beta}_{(l)}^{(t)}, (\mathbf{b}_1)_{(l)}^{(t)}, \dots, (\mathbf{b}_{k-1})_{(l)}^{(t)}, (\mathbf{b}_{k+1})_{(l)}^{(t-1)}, \dots, (\mathbf{b}_{K+1})_{(l)}^{(t-1)}, \\
        (\boldsymbol{\sigma}^2_{k})_{(l)}^{(t-1)}, (\mathbf{\Sigma}_Z)_{(l)}^{(t-1)}, (\boldsymbol{\sigma}^2_e)_{(l)}^{(t-1)}, \mathbf{X}_{(l)}, \mathbf{y}_{(l)}) \),
        \item[(2.3)] sample for \(k = 1, \dots, K\): \\
        \( (\mathbf{\sigma}^2_{k})_{(l)}^{(t)} \sim \\
        p_{(l)}(\mathbf{\sigma}^2_{k} | \boldsymbol{\beta}_{(l)}^{(t)}, \mathbf{b}_{(l)}^{(t)}, (\mathbf{\sigma}^2_1)_{(l)}^{(t)}, \dots, (\sigma^2_{k-1})_{(l)}^{(t)}, (\sigma^2_{k+1})_{(l)}^{(t-1)}, \dots, (\sigma^2_{K})_{(l)}^{(t-1)}, \\ 
        (\mathbf{\Sigma}_Z)_{(l)}^{(t-1)}, (\boldsymbol{\sigma}^2_e)_{(l)}^{(t-1)}, \mathbf{X}_{(l)}, \mathbf{y}_{(l)}) \),
        \item[(2.4)] sample \( (\boldsymbol{\mathbf{\Sigma}}_Z)_{(l)}^{(t)} \sim p_{(l)}(\boldsymbol{\mathbf{\Sigma}}_Z | \boldsymbol{\beta}_{(l)}^{(t)}, \mathbf{b}_{(l)}^{(t)}, (\boldsymbol{\sigma}^2_{k})_{(l)}^{(t)}, (\boldsymbol{\sigma}^2_e)_{(l)}^{(t-1)}, \mathbf{X}_{(l)}, \mathbf{y}_{(l)}) \),
        \item[(2.5)] sample for \(e = 1, \dots, E\): \\
        \( (\mathbf{\sigma}^2_{e})_{(l)}^{(t)} \sim \\
        p_{(l)}(\mathbf{\sigma}^2_{e} | \boldsymbol{\beta}_{(l)}^{(t)}, \mathbf{b}_{(l)}^{(t)}, (\boldsymbol{\sigma}^2_k)_{(l)}^{(t)}, (\mathbf{\Sigma}_Z)_{(l)}^{(t)}, \\
        (\mathbf{\sigma}^2_1)_{(l)}^{(t)}, \dots, (\sigma^2_{e-1})_{(l)}^{(t)}, (\sigma^2_{e+1})_{(l)}^{(t-1)}, \dots, (\sigma^2_{E})_{(l)}^{(t-1)}, \mathbf{X}_{(l)}, \mathbf{y}_{(l)}) \),
    \end{itemize}      
    where \(p_{(l)}(\boldsymbol{\beta} | \cdot)\) of (2.1) involves for all windows \(l\) by default of the rstan package an uninformative improper uniform prior such that only the likelihood information of the current multi-year dataset is taken into account, \\
    and \(p_{(l)}(\mathbf{b}_{k} | \cdot)\) of (2.2) involves for \(l = 1\) the prior \(p_{(1)}(\mathbf{b}_k | (\sigma^2_k)^{(t-1)}_{(1)}) =\mathcal{N} \left( \mathbf{0}, (\sigma^2_k)^{(t-1)}_{(1)}\mathbf{I}_{Q_k}\right)\) \\
    or \(p_{(1)}(\mathbf{b}_{K+1} | (\mathbf{\Sigma}_Z)^{(t-1)}_{(1)}) = \mathcal{N} \left( \mathbf{0}, \text{block-diag}[(\mathbf{\Sigma}_Z)^{(t-1)}_{(1)}, \dots, (\mathbf{\Sigma}_Z)^{(t-1)}_{(1)}]\right)\) \\ 
    and \(\forall \ l \neq 1\) the updated prior \(p_{(l)}(\mathbf{b}_k | (\sigma^2_k)^{(t-1)}_{(l)}) = \mathcal{N} \left( \mathbf{0}, (\sigma^2_k)^{(t-1)}_{(l)} \mathbf{I}_{Q_k}\right)\) \\
    or \(p_{(l)}(\mathbf{b}_{K+1} | (\mathbf{\Sigma}_{Z})^{(t-1)}_{(l)}) = \mathcal{N} \left( \mathbf{0}, \text{block-diag}[(\mathbf{\Sigma}_Z)^{(t-1)}_{(l)}, \dots, (\mathbf{\Sigma}_Z)^{(t-1)}_{(l)}]\right)\) \\
    where \(p_{(l)}(\mathbf{b}_k | (\sigma^2_k)^{(t-1)}_{(l)})\) and \(p_{(l)}(\mathbf{b}_{K+1} | (\mathbf{\Sigma}_{Z})^{(t-1)}_{(l)})\) involving \(\mathbf{X}_{(l-1)}, \mathbf{y}_{(l-1)}\) implicitly, \\
    and \(p_{(l)}(\mathbf{\sigma}^2_{k} | \cdot)\) of (2.3) involves for \(l = 1\) the prior \(p_{(1)}(\sigma^2_k) = \text{Inv-Gamma} \left( u, 1 \right)\) with \(u\) in the range of 5 up to 60 dependent on the specific variance component and \(\forall \ l \neq 1\) the updated prior \(p_{(l)}(\sigma^2_k | \mathbf{X}_{(l-1)}, \mathbf{y}_{(l-1)}) = \text{Inv-Gamma} \left( (\widehat{\alpha}_{k})^{(\text{MLE})}_{(l-1)}, (\widehat{\beta}_{k})^{(\text{MLE})}_{(l-1)} \right)\), \\
    and \(p_{(l)}(\boldsymbol{\mathbf{\Sigma}}_Z | \cdot)\) of (2.4) involves for \(l = 1 \) the prior  \(p_{(1)}(\mathbf{\Sigma}_Z) = \text{Inv-Wishart}(20, \mathbf{I}_Z)\) and \(\forall \ l \neq 1\) the updated prior \(p_{(l)}(\mathbf{\Sigma}_Z | \mathbf{X}_{(l-1)}, \mathbf{y}_{(l-1)}) = \text{Inv-Wishart}\left((\widehat{\nu})^{(\text{MLE})}_{(l-1)}, (\widehat{\mathbf{S}})^{(\text{MLE})}_{(l-1)} \right)\), \\
    and \( p_{(l)}(\mathbf{\sigma}^2_{e} | \cdot)\) of (2.5) involves for all \(l = 1\) the prior \(p_{(1)}(\sigma^2_e) =\text{Inv-Gamma}(10, 1)\) and \(\forall \ l \neq 1\) the updated prior \(p_{(l)}(\sigma^2_e | \mathbf{X}_{(l-1)}, \mathbf{y}_{(l-1)}) = \text{Inv-Gamma}\left((\widehat{\alpha}_{e})^{(\text{MLE})}_{(l-1)}, (\widehat{\beta}_{e})^{(\text{MLE})}_{(l-1)}\right)\).
    \item[(3)] Discard the \(B\) burn-in draws.
    \item[(4)] Compute and store \((\widehat{\alpha}_{k})^{(\text{MLE})}_{(l)}, (\widehat{\beta}_{k})^{(\text{MLE})}_{(l)},(\widehat{\nu})^{(\text{MLE})}_{(l)}, (\widehat{\mathbf{S}})^{(\text{MLE})}_{(l)},(\widehat{\alpha}_{e})^{(\text{MLE})}_{(l)}, (\widehat{\beta}_{e})^{(\text{MLE})}_{(l)}\) \(\forall \ k\) and \(\forall \ e\) as described in Section \ref{app:cond:MCMC:MLE} via ML estimation based on their current posterior sample.
\end{enumerate}

\subsection{Mathematical background of A-optimal design for sub-divided target population of environments}
\label{app:opti:design}
\textcite{Prus2021} specified an A-optimality design criterion in our specific data and model context. The idea was to derive a criterion from the MSE matrix trace of the linear contrasts of genotype BLUPs that is a function of both the experimental MET design and the genotype-dependent random-effects variance components of interest. Without considering multiple years in the MET, the following design criterion was derived.
\begin{equation} \label{A:design:without:year}
\Phi_A(\boldsymbol{\xi_a}) = \text{tr} \left( \left( \mathbf{M}(\boldsymbol{\xi_a}) + \Delta^{-1} \right)^{-1} \right),
\end{equation}
where $\mathbf{M}(\boldsymbol{\xi_a}) = \text{diag}(w_1, \dots, w_Z)$ with $\sum^{Z}_{z = 1} w_Z = 1$ and $w_Z \geq 0$ is the information matrix of the approximate design, and $\Delta$ is an adjusted covariance matrix dependent on the variance components in the underlying LMM. The approximate design refers to weights \(w_Z\) that add up to one for the different zones. On the other hand, the exact design \(\boldsymbol{\xi_{ex}}\) refers to integer numbers of trials \(J_z\) that should be allocated to each zone, as directly derived from the approximate design, with \(\sum^{Z}_{z = 1} J_z = J\) being the total number of trials in the experiment. For the exact design, it holds that \(w_Z = J_z/J\). The allocation strategy depends on the chosen covariance structure of the genotype-zone random effect. \\
\textcite{Prus2024} extended their work from before, especially considering the multi-year structure in MET. An extended A-optimal design is required, incorporating the effects of year and genotype-year random effects. The extended criterion, neglecting constants that do not influence the design, accounts for multiple years in the experimental design by modifying the A-optimality design criterion as follows:
\begin{equation} \label{A:design:with:year}
\Phi^{\star}_A(\boldsymbol{\xi_a}) = \text{tr} \left[ \left( \mathbf{M}(\boldsymbol{\xi_a}) + \mathbf{Q}^{-1} \right)^{-1} \mathbf{B}^{-1} \mathbf{K} \mathbf{K} \mathbf{B}^{-1} \right],
\end{equation}
where $\mathbf{M}(\boldsymbol{\xi_a})$ is again the information matrix of the approximate design, containing information about trial allocations across zones. $\mathbf{K}$ is a matrix describing the genotype-specific variance-covariance structure of zones. Note that $\mathbf{K}$ is defined as $\mathbf{V}$ in the work of \cite{Prus2024}. $\mathbf{B}$ is an adjusted variance-covariance matrix incorporating additional to $\mathbf{K}$ genotype-specific random effect variances across years scaled down by the total number of years in the design. $\mathbf{Q}$ is equal to $\frac{J}{\kappa} \mathbf{B}$, where $J$ is the total number of trial locations, and $\kappa$ is a scaling constant depending on the total number of years as well as genotype-specific random effect variances across locations and the residual error variance scaled down by the number of replicates in the balanced dataset. Note that $\kappa$ is defined as $c$ in the work of \cite{Prus2024}.\\
The criteria $\Phi_A(\boldsymbol{\xi})$ and $\Phi^{\star}_A(\boldsymbol{\xi})$ are explicitly derived for the data structure provided by \textcite{Kleinknecht2013} and its statistical models as specified by \textcite{Prus2021} for equation \eqref{A:design:without:year} and by \textcite{Prus2024} for equation \eqref{A:design:with:year}. These criteria cannot be understood as general results usable for any MET data and need to be uniquely derived in a certain LMM. Fortunately, the data structure of Section \ref{sec:data} in this study is very similar to the data of \textcite{Kleinknecht2013} with different number of locations, zones, and years. The statistical model used, as specified in Section \ref{sec:methods:MM:specification}, is a modification of the model (2) specified in the work of \textcite{Prus2024}, where environment-specific residual error variances are considered in the model. To match $\Phi^{\star}_A(\boldsymbol{\xi})$ by \textcite{Prus2024} in this study, the environmental mean across all environment-specific variance components env\_mean\_var\_resid was extracted and used in the optimal design calculation. The uncertainty captured in this residual environmental mean estimator is implicitly taken into account, considering our Bayesian reformulation of the frequentist LMM provided in Section \ref{sec:bayesmod:bayes:cond}.

\subsubsection{Efficiency}
\label{method:Eff}
The approximate optimal design efficiencies \( Eff_{a}\) are used to evaluate the performance of experimental designs (\cite{Prus2021, Prus2024}). The efficiency \(Eff_{a}\) measures the relative performance of an optimal approximate design \( \boldsymbol{\xi_a^{\star}} \) compared to the completely balanced design \( \boldsymbol{\xi_P} \). It is mathematically defined as
\begin{equation}
Eff_{a} = \frac{\Phi_A(\boldsymbol{\xi_a^{\star}})}{\Phi_A(\boldsymbol{\xi_P})}.
\end{equation}
The efficiencies \(Eff_{a}\) range between 0 and 1, with values closer to 1 indicating that the optimal design $\boldsymbol{\xi_a^{\star}}$ is as good as the completely balanced design $\boldsymbol{\xi_P}$. However, if the efficiency \(Eff_{a}\) remains below 1, this means that the optimal design $\boldsymbol{\xi_a^{\star}}$ is better than the completely balanced design $\boldsymbol{\xi_P}$.

\subsubsection{Mean squared error matrix trace}
\label{method:MSE:Trace}
The trace of the MSE matrix, denoted as \( MSE_{\text{Tr}} \), serves to evaluate the accuracy of predictions of pairwise linear contrasts of genotype BLUPs under a given design. Its mathematical formula is very similar to the A-criterion of equation \eqref{A:design:with:year}, additionally taking variance component dependent factors into account that are irrelevant for the optimal design optimization itself. The exact formula can be found in equation (16) of the work of \textcite{Prus2024}. Lower values of \( MSE_{\text{Tr}} \) indicate designs that yield more accurate predictions. 

\end{document}